\documentclass[fleqn,usenatbib]{mnras}

\usepackage{newtxtext,newtxmath}
\usepackage{longtable}
\usepackage[T1]{fontenc}
\usepackage{ae,aecompl}

\DeclareRobustCommand{\VAN}[3]{#2}
\let\VANthebibliography\thebibliography
\def\thebibliography{\DeclareRobustCommand{\VAN}[3]{##3}\VANthebibliography}

\usepackage{graphicx}	% Including figure files
\usepackage{amsmath}	% Advanced maths commands
\usepackage{siunitx}
\usepackage{xspace}
\usepackage{orcidlink}
\usepackage{xcolor}

\newcommand{\lo}{L_{\rm UV}}
\newcommand{\lx}{L_{\rm X}}
\newcommand{\nh}{N_{\rm H}}
\newcommand{\nhz}{N_{\rm H}(z)}

\newcommand{\sbline}{$\rm erg~s^{-1}~cm^{-2}~arcsec^{-2}$}
\newcommand{\qsoone}{J2142$-$4420\xspace}
\newcommand{\qsotwo}{J2142$-$4419\xspace}
\newcommand{\qsothree}{LBQS 2138$-$4427\xspace}
    \newcommand{\xmm}{\emph{XMM--Newton}\xspace}
    \newcommand{\swift}{\emph{Swift}\xspace}
    
    \newcommand{\chandra}{\emph{Chandra}\xspace}
    \newcommand{\nustar}{\emph{NuSTAR}\xspace}
    
    \newcommand{\spitzer}{\emph{Spitzer}\xspace}
    
    \newcommand{\hst}{\emph{HST}\xspace}
    \renewcommand{\ang}{\AA\xspace}

\newcommand{\qsfit}{\texttt{QSFit}\xspace}
\newcommand{\xspec}{\textsc{xspec}\xspace}
\newcommand{\sas}{\textsc{sas}\xspace}
\newcommand{\arfgen}{\textsc{arfgen}\xspace}
\newcommand{\rmfgen}{\textsc{rmfgen}\xspace}

\newcommand{\epicspeccombine}{\textsc{epicspeccombine}\xspace}

\DeclareRobustCommand{\ion}[2]{%
\relax\ifmmode
\ifx\testbx\f@series
{\mathbf{#1\,\mathsc{#2}}}\else
{\mathrm{#1\,\mathsc{#2}}}\fi
\else\textup{#1\,{\mdseries\textsc{#2}}}%
\fi}

\newcommand{\lya}{Ly$\alpha$}
\newcommand{\gx}{\Gamma_{\rm X}}
\newcommand{\kms}{km s$^{-1}$}
\newcommand{\ebv}{E(B-V)}

\newcommand{\rev}[1]{{ #1}}
\newcommand{\revs}[1]{{ #1}}

\title[MUDF. IV. X-ray properties of the quasar pair]{The MUSE Ultra Deep Field (MUDF). IV. A pair of X-ray weak quasars at the heart of two extended \lya\ nebulae}

\author[E. Lusso et al.]{Elisabeta Lusso\orcidlink{0000-0003-0083-1157}$^{1,2}$\thanks{E-mail: elisabeta.lusso@unifi.it},
Emanuele Nardini\orcidlink{0000-0001-9226-8992}$^{2}$,
Michele Fumagalli\orcidlink{0000-0001-6676-3842}$^{3,4}$,
Matteo Fossati\orcidlink{0000-0002-9043-8764}$^{3,5}$,
Fabrizio Arrigoni \newauthor Battaia\orcidlink{0000-0002-4770-6137}$^{6}$,
Mitchell Revalski\orcidlink{0000-0002-4917-7873}$^{7}$,
Marc Rafelski\orcidlink{0000-0002-9946-4731}$^{7,8}$,
Valentina D'Odorico\orcidlink{0000-0003-3693-3091}$^{9,10,11}$,
Celine Peroux\orcidlink{0000-0002-4288-599X}$^{12,13}$, \newauthor
Stefano Cristiani\orcidlink{0000-0002-2115-5234}$^{4,11,14}$,
Pratika Dayal\orcidlink{0000-0001-8460-1564}$^{15}$,
Francesco Haardt\orcidlink{0000-0003-3291-3704}$^{16}$,
Emma K. Lofthouse\orcidlink{0000-0002-1209-9680}$^{3,5}$
\\
$^{1}$Dipartimento di Fisica e Astronomia, Universit\`a di Firenze, via G. Sansone 1, 50019 Sesto Fiorentino, Firenze, Italy\\
$^{2}$INAF -- Osservatorio Astrofisico di Arcetri, Largo Enrico Fermi 5, I-50125 Firenze, Italy\\
$^{3}$Dipartimento di Fisica G. Occhialini, Universit\`a degli Studi di Milano-Bicocca, Piazza della Scienza 3, 20126 Milano, Italy\\
$^{4}$INAF - Osservatorio Astronomico di Trieste, via G. B. Tiepolo 11, 34143 Trieste, Italy\\
$^{5}$INAF - Osservatorio Astronomico di Brera, via Bianchi 46, 23087 Merate (LC), Italy\\
$^{6}$Max-Planck-Institut fur Astrophysik, Karl-Schwarzschild-Str 1, D-85748 Garching bei M\"unchen, Germany\\
$^{7}$Space Telescope Science Institute, 3700 San Martin Drive, Baltimore, MD 21218, USA\\
$^{8}$Department of Physics and Astronomy, Johns Hopkins University, Baltimore, MD 21218, USA\\
$^{9}$INAF--Osservatorio Astronomico di Trieste, Via G.B. Tiepolo, 11, I-34143 Trieste, Italy \\
$^{10}$Scuola Normale Superiore, P.zza dei Cavalieri, I-56126 Pisa, Italy\\
$^{11}$IFPU--Institute for Fundamental Physics of the Universe, via Beirut 2, I-34151 Trieste, Italy\\
$^{12}$European Southern Observatory, Karl-Schwarzschildstrasse 2, D-85748 Garching bei M{\"u}nchen, Germany\\
$^{13}$Aix Marseille Universit\'e, CNRS, LAM (Laboratoire d'Astrophysique de Marseille) UMR 7326, 13388, Marseille, France \\
$^{14}$INFN–National Institute for Nuclear Physics, via Valerio 2, I-34127 Trieste, Italy\\
$^{15}$Kapteyn Astronomical Institute, Rijksuniversiteit Groningen, Landleven 12, Groningen, 9717 AD, The Netherlands\\
$^{16}$DiSAT, Università degli Studi dell’Insubria, via Valleggio 11, I-22100 Como, Italy
}

\date{\today}
\pubyear{2023}

\begin{document}
\label{firstpage}
\pagerange{\pageref{firstpage}--\pageref{lastpage}}
\maketitle

% Abstract of the paper
% 250 words, no refs
\begin{abstract}
We present the results obtained from follow-up observations of the MUSE Ultra Deep Field (MUDF) at X-ray energies with \xmm. The MUDF is centred on a unique field with two bright, physically associated quasars at $z\simeq3.23$, separated by $\sim$500 kpc in projection. 
Both quasars are embedded within extended Ly$\alpha$ nebulae ($\gtrsim 100~\rm kpc$ at a surface brightness flux level of $\approx 6\times 10^{-19}$~\sbline), whose elongated morphology is suggestive of an extended filament connecting the quasar haloes. %Additionally, both quasars display associated absorption features in their ultraviolet spectra, which reveal the presence of enriched material. 
The new X-ray observations presented here allow us to characterise the physical properties (e.g. X-ray slope, luminosities, gas column densities) in the innermost region of the MUDF quasars. 
We find that both quasars are X-ray underluminous compared to objects at similar ultraviolet luminosities. Based on our X-ray spectral analysis, absorbing columns of $\nhz\gtrsim$\,10$^{23}$ cm$^{-2}$ appear unlikely, therefore such a weakness is possibly intrinsic. %At the same time, these quasars present an \rev{``excess'' of \ion{C}{iv} line emission with respect to the one of typical sources at similar X-ray luminosities that follows the X-ray to \ion{C}{iv} correlation}.  
When also including literature data, we do not observe any detectable trend between the area of the nebulae and nuclear luminosities at both the rest-frame 2 keV and 2500 \AA. The area is also not correlated with the X-ray photon index nor with the integrated band flux in the hard band (2--10 keV). We also do not find any trend between the extended Ly$\alpha$ emission of the nebulae and the nuclear X-ray luminosity. \rev{Finally, the properties of the MUDF quasars' nebulae are consistent with the observed relation between the \lya\ integrated luminosity of the nebulae and their area.} Our results suggest that the quasar ionization power is not a strong driver of the morphology and size of the nebulae. % final extent of the nebulae and their morphology is mainly driven by the environment, rather than the quasar power.
\end{abstract}

% Select between one and six entries from the list of approved keywords.
% Don't make up new ones.

%%%%%%%%%%%%%%%%%%%%%%%%%%%%%%%%%%%%%%%%%%%%%%%%%%

%%%%%%%%%%%%%%%%% BODY OF PAPER %%%%%%%%%%%%%%%%%%

\section{Introduction}
The current cosmological concordance model ($\Lambda$CDM) predicts that galaxies form in overdensities at the intersection of filaments, which compose the cosmic web  (e.g. \citealt{bond1996,2018PhR...780....1D}). This prediction is supported indirectly by observations of the large scale structures traced with galaxy surveys \citep[e.g.][]{reid2012,anderson2014,wisotzki2018,malavasi2020} and by studies of the \lya\ forest in absorption \citep[e.g.][]{mcdonald2005}. 
A valuable technique to map the gas distribution in the circumgalactic medium (CGM) has been through the direct imaging of the fluorescent \lya\ line in emission around bright quasars (e.g. \citealt{cantalupo2014,hennawi2015,borisova2016,2018MNRAS.473.3907A,arrigoni2019a,arrigoni2019b,farina2019,cai2019,osullivan2020,fossati2021}) and galaxies \citep{Leclercq2017,wisotzki2018}, where the integral field spectrographs positioned at the largest observing facilities, such as the Multi-Unit Spectroscopic Explorer \citep[MUSE;][]{bacon2010} at the Very Large Telescope and the Keck Cosmic Web Imager \citep[KCWI;][]{morrisey2018} at the Keck telescope, have been key to significantly improve the detection of this extended \lya\ emission \citep[see also e.g.][]{umehata2019,bacon2021,Kusakabe2022,ln2022}. 

% some more discussion of nebulae with X-ray follow-up
\rev{Understanding the ionising source of extended \lya\ nebulae requires multi-wavelength data as obscured AGN, for instance, can be missed at UV energies \citep[e.g.][]{geach2009}. Indeed, rest-frame UV-based surveys are blind to dust obscured quasars, whilst X-rays (together with radio wavelengths) can help to constrain the presence (or not) of possible sources associated to extended \lya\ emission \citep[see][for a review on the topic]{cantalupo2017}.}
  
\rev{\lya\ nebulae have not yet been systematically targeted with deep X-ray observations, as the presence of a hard ionising source, the quasar, is usually given for granted. However, X-ray explorations have been conducted to understand the nature of the so-called \lya\ blobs \citep[LABs;][]{Steidel2001} in which the identification of the powering sources turned out to be difficult.
To our knowledge, the only field with deep X-ray coverage in the literature is the Small Selected Area 22h (SSA22) field \citep{lilly1991,steidel1998}. %SSA22 has been extensively studied at several wavelengths and revealed the presence of a bright submillimeter source with a bolometric luminosity in excess of $10^{13}L_\odot$ \citep[e.g.][]{chapman2001,geach2005}. 
SSA22 is an overdense region at $z=3.1$ known to host several LABs \citep[e.g.,][]{Matsuda2004}. The field has been extensively studied at several wavelengths. First evidence of embedded powering sources in the LABs came from the detection of bright submillimetre objects with a bolometric luminosity in excess of $10^{12}$~L$_{\odot}$ \citep[e.g.][]{chapman2001,geach2005}. 
No evidence from deep \chandra\ X-ray observations of a clear X-ray counterpart was found \citep{chapman2004}, yet the possibility of a luminous but heavily obscured AGN along our line of sight was not excluded. By analysing the same data, \citet{bzs2004} discovered a hard-band X-ray source in a second giant \lya\ emission nebula in the SSA22 region. They confirmed that the nebula with significant submillimetre output is undetected at the X-ray energies, whilst the other is a factor of 2--10 times less bright in the submillimetric but appears to contain a highly obscured AGN. Neither of these two \lya\ nebulae were associated with conspicuous radio emission. With even deeper \chandra\ observations (400 ks), \citet{lehmer09apj} found a total of five X-ray detected sources in 158 \lya\ emitters in the SSA22 field, implying a larger fraction of AGN activity than in lower density environments (see also \citealt{monson23}).} 

\rev{\citet{nilsson2006} published the discovery of a \lya\ nebula at $z \simeq3.157$ in the Great Observatories Origins Deep Survey (GOODS) South field, which is rich in multiwavelength data from the X-rays with \chandra\ to the infrared with \spitzer. Intriguingly, unlike other known \lya\ extended nebulae, the deep GOODS imaging of the nebulae displayed no associated continuum counterparts in any band, from the X-rays to the infrared, suggesting that the most probable origin of the extended \lya\ emission could be cold accretion onto a dark matter halo rather than an AGN. This scenario, however, was later disproved by the detection of six continuum sources associated with the nebula and a nearby obscured AGN \citep{prescott15b}, which actually turns out to be located at the center of a larger \lya\ structure \citep{sanderson21}.}

\rev{LABs are also found at much lower redshifts. \citet{Taiki2017} presented \nustar\ hard X-ray (3--30 keV) observations, complemented with \chandra\ and \swift\ data, of the two X-ray brightest sources at $z \simeq 0.3$ that show extended \lya\ emission, SDSS J011341.11+010608.5 and SDSS J115544.59$-$014739.9 \citep{schirmer2016}. \nustar data revealed the presence of bright X-ray emission in both sources, with 2--10 keV luminosity $(0.5-3)\times10^{44}$ erg s$^{-1}$ and moderate obscuration ($\nh \sim 0.6-5 \times 10^{23}$ cm$^{-2}$).}

%MUDF
\revs{In this framework, our group designed an observational campaign to acquire very deep observations with MUSE in a $\approx$$1.2\times 1.4$ arcmin$^2$ region centred at RA=21:42:24 and Dec=$-$44:19:48 (hereafter the MUSE Ultra Deep Field or MUDF). The MUDF hosts several astrophysical structures at different redshifts, including two physically associated quasars at $z\approx 3.23$, %(\qsoone\ and \qsotwo) 
J214225.78$-$442018.3 and J214222.17$-$441929.8 (hereafter \qsoone and \qsotwo, respectively), with a projected separation of $\approx 1$ arcmin (or $\approx 500~\rm kpc$ at $z\approx 3$). Another quasar with similar redshift lies at $\approx$8 arcmin separation (\qsothree), making this system a quasar triplet \citep{fh1993}. 
%Upon its completion, this programme will acquire $\approx200$ h of MUSE data (ESO PID 1100.A$-$0528, see \citealt{fossati2019} for details about the survey design and the details of the MUSE observations and data reduction). 
In the first paper of this series, \citet{lusso2019b} studied the morphology of the extended \lya\ nebulae surrounding the quasars, observing that the ionized gas is elongated along the line connecting the two sources. In the second paper, \cite{fossati2019} studied of the galaxy environment and gaseous properties of the seven galaxy groups detected at redshifts $0.5 < z < 1.5$ with halo mass in the interval $\log (M_{\rm h}/{M_\odot}) \simeq 11 - 13.5$. The absorption strength associated with these groups is higher to that of near isolated galaxies at similar mass and impact parameters. Additionally, no evidence was found for widespread cool gas that could be associated to a diffuse intra-group medium. In the third paper, \cite{revalski2023} utilised the extensive wavelength coverage of MUSE and WFC3 to measure spectroscopic redshifts for 419 sources down to galaxy stellar masses of $\log(M/M_\odot) \simeq 7$ at $z \simeq 1-2$, and publicly released the calibrated Hubble Space Telescope (\hst) observations, catalogues, and associated data products\footnote{\url{https://archive.stsci.edu/hlsp/mudf}}.}

In this paper, we present the \xmm\ observation of the MUDF, which provides the first view at high energies of the assembly of a potentially massive $z\simeq3.2$ overdensity in this field. Our main goal is to characterise the physical properties (e.g. X-ray slope, luminosities, gas column densities) in the innermost region of quasars with physically associated and extended \lya\ nebulae.

The paper is structured as follows: in Section~\ref{obs and data red} we present the MUDF and \xmm\ observations and data reduction, whilst the X-ray and the UV spectral analyses are discussed in Section~\ref{x-ray analysis} and Section~\ref{Ultraviolet spectral analysis}, respectively.  Section~\ref{results and discussion} is devoted to the presentation of the results and their discussion, with conclusions drawn in Section~\ref{conclusions}.
Whenever luminosity values are reported, we have assumed a standard flat $\Lambda$CDM cosmology with $\Omega_{\rm M}=0.3$ and $H_0=70$ km s$^{-1}$ Mpc$^{-1}$.

% --------
\begin{figure*}
\centering\includegraphics[width=\hsize]{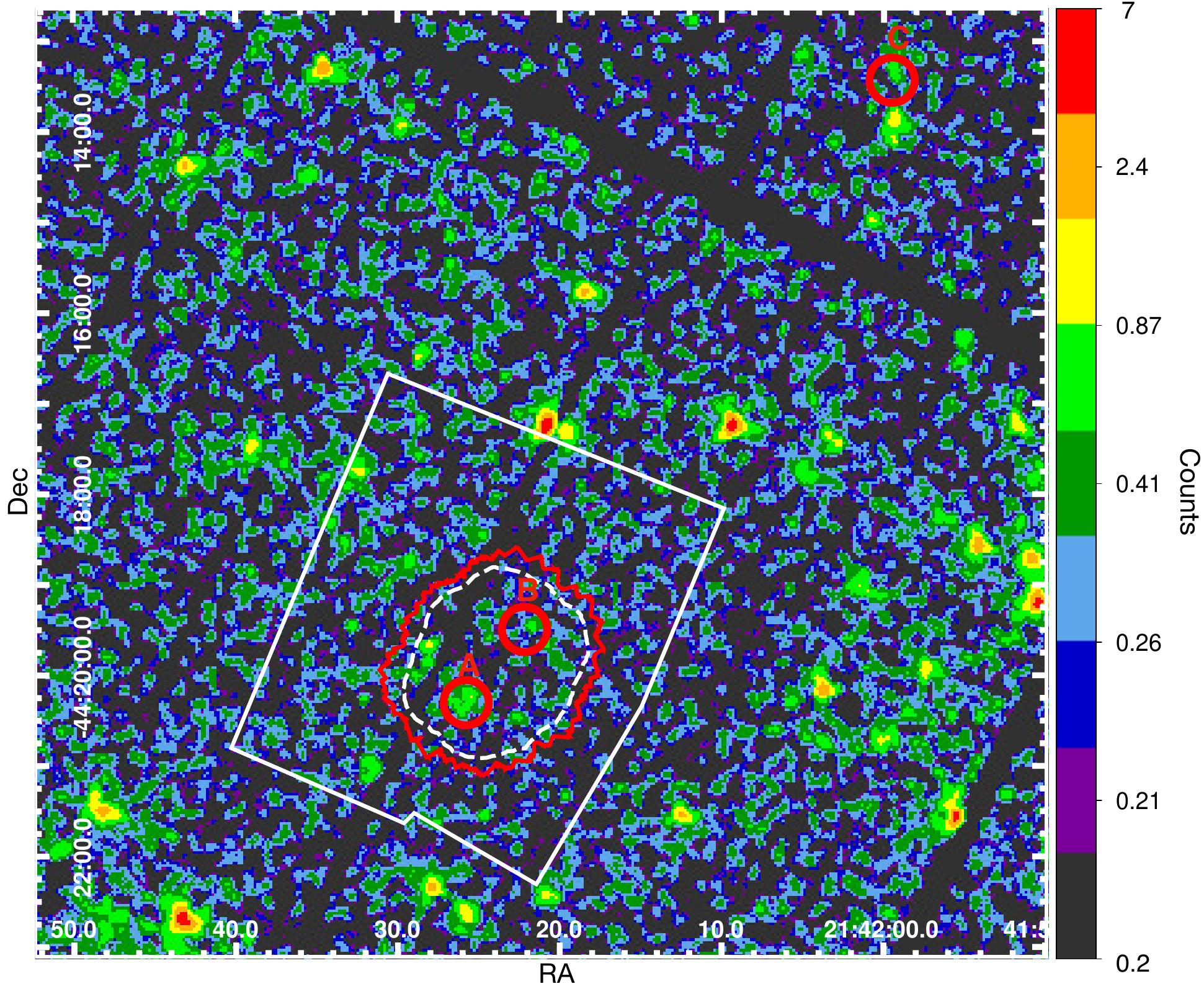}
    \caption{\xmm MOS2 (Metal Oxide Semiconductor) image at 0.5--5 keV. The area marked with the white solid line represents the F140W/\hst\ coverage \citep{revalski2023}. The solid red line represents the MUDF covered with MUSE, whilst the white dashed lines marks the region with at least 4 hours of exposure with MUSE. The red circles mark the locations of the primary quasars and have radii of $\sim15^{\prime\prime}$. The image is spatially rebinned and smoothed through a Gaussian function. Labels A, B, and C as in Table~\ref{tab:sum}. \rev{The (cleaned) \xmm images for the pn, MOS1, and MOS2 CCDs that cover the 0.2--12 keV energy band (i.e., EPIC band 8) of the MUDF can be accessed through MAST at \url{https://archive.stsci.edu/hlsp/mudf}.}}
    \label{fig:xmmfield}
\end{figure*}
% --------
\section{Observations and data reduction}
\label{obs and data red}

\subsection{The MUSE Ultra Deep Field data}
The MUDF is covered by $\approx$140 h of MUSE data (ESO PID 1100.A$-$0528, see \citealt[][]{fossati2019}, for details about the survey design, the MUSE observations and data reduction; Fossati et al. in preparation for the presentation of the final dataset), with $\approx$100 h in the centremost region.
This programme is complemented by deep high resolution spectroscopy of the quasars with UVES (Ultraviolet Visual Echelle Spectrograph; \citealt{dekker2000}) at the VLT (ESO PIDs 65.O$-$0299, 68.A$-$0216, 69.A$-$0204, 102.A$-$0194; \citealt{dodorico2002}), and by a very deep spectroscopic survey (90 orbits in a single field) in the near-infrared with the Wide Field Camera 3 (WFC3) instrument on board the \hst\ \citep{revalski2023}, together with deep eight-orbit near UV imaging (program IDs: 15637, PI: M. Rafelski; and 15968, PI: M. Fossati). 
\rev{Specifically, the MUDF has been observed with the WFC3/IR %onboard \hst\ for 90 orbits 
with the G141 grism and the F140W filter, %(program ID 15637, PIs: M. Rafelski and M. Fumagalli). Spectroscopic observations with the G141 grism 
which covers a spectral range of 10750--17000 \AA, and thus extends the MUSE spectroscopic data in the rest-frame UV, with a resolving power of $R\simeq150$ at 14000 \AA. We refer to \citet[][]{revalski2023} for details on the program design and acquisition of the \hst\ observations.}

\subsection{The X-ray data}
\xmm\ targeted the MUDF for a full orbit (revolution 3731, duration 139 ks; PI: E. Lusso) on 2020 April 22, with the three EPIC (European Photon Imaging Camera) cameras \citep{struder+01,turner+01} operating in Full Frame mode with thin optical filter. The event files were reprocessed with the Science Analysis System (\textsc{sas}) v19.1.0, following the standard procedures and using the latest calibration files. The final part of the observation was affected by background flares, hence the good time intervals have been created by imposing a count rate threshold for high-energy single-pixel events of 0.4 s$^{-1}$ (10--12 keV) and 0.35 s$^{-1}$ ($>$10 keV) over the whole pn and MOS detectors\footnote{\url{https://www.cosmos.esa.int/web/xmm-newton/technical-details-epic}}, respectively. After the dead-time correction and background filtering, the net exposures are 94.8 ks for pn, 128.5 ks for MOS1, and 128.6 ks for MOS2. 

The source spectra were extracted from circular regions centred at the nominal optical coordinates of each target of interest, with radii of 15$\arcsec$ for \qsoone\ (and \qsothree) and 12$\arcsec$ for \qsotwo, corresponding to an encircled energy fraction of $\lesssim$70\,\%. Although a non negligible fraction of counts might be lost,\footnote{Note that this `aperture loss' is corrected for at the spectral analysis stage by the ancillary response files, which store all the \textit{geometrical} information relative to the source extraction position on each detector (e.g. effective area, quantum efficiency, vignetting).} such apertures are required due to the presence of nearby point-like or diffuse emission structures that would otherwise contaminate the source spectra (see Figure~\ref{fig:xmmfield} and Section~\ref{xraysourcedet}). The background was evaluated on nearby regions free of excess emission, with radii of 40$\arcsec$--60$\arcsec$ depending on the target and detector. The total number of collected net counts is provided for each source in Table~\ref{tab:sum}. We note that \qsothree\ falls outside the MOS1 field of view, while in the pn image \qsoone\ partly falls on the gap between two adjacent chips, thus reducing the actual effective area. The appropriate response files were generated with the \sas tasks \rmfgen and \arfgen.  

\begin{figure}
\centering\includegraphics[width=\hsize]{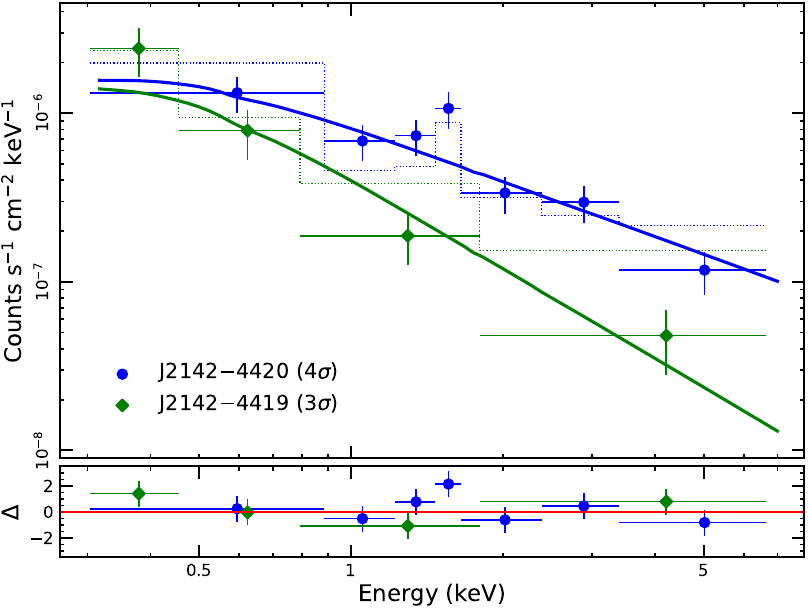}
    \caption{Combined \xmm EPIC spectra of \qsoone (blue dots) and \qsotwo (green diamonds), binned for visual clarity to a 4$\sigma$ and 3$\sigma$ significance, respectively. The best power-law fits are represented by the solid curves, while the dotted lines indicate the background levels. Residuals are computed as $\Delta$\,$=$\,(data$-$model)/error and are shown in the bottom panel. Note that the line-like excess around 1.6 keV in the spectrum of \qsoone, although corresponding to the Fe\,K band in the rest frame, is most likely an instrumental artefact, as discussed in the text. The red line refers to $\Delta=0$.}
    \label{fig:xspecs}
\end{figure}

The spectral analysis was performed over the 0.3--7 keV band with the \xspec v12.12.0 fitting package \citep{arnaud1996}. The spectra are fully background-dominated at higher energies, and changing the upper end of the fitting range has no effect on the results. Given the limited statistics, the spectra were binned to ensure a minimum of one count per energy channel and \rev{a modified version of the Cash statistic was adopted \citep[\textit{cstat}, or W-statistic in \xspec;][]{cstat,kaastra17}, as appropriate for source and background data in the Poissonian regime}. The uncertainties we provide on the model parameters, including fluxes, correspond to $\Delta C = 1$, unless stated otherwise. The fits were simultaneously carried out on the individual spectra from the three EPIC detectors. Yet, for visual purposes, we also generated a merged EPIC spectrum with the \sas task \epicspeccombine (see Figure~\ref{fig:xspecs}).

% --------
% --------
% -----------------
\begin{table*}
\centering
\caption{Summary of the properties of the MUDF quasars.} \label{tab:sum}
\begin{tabular}{lccccccccc}
\hline
\hline
Object & $z_{\rm sys}$ & $(\nu L_{\nu})_{\rm UV}^{\rm a}$ & $(\nu L_{\nu})_{\rm X}^{\rm b}$ & $(\nu L_{\nu})_{\rm CIV}^{\rm c}$ & Cts$^{\rm d}$ & $\gx^{\rm e}$ & $(\nu F_{\nu})_{\rm 2\,{\rm keV}}^{\rm f}$ & $C/\nu^{\rm g}$ & $(\nu F_{\nu})_{\rm 2\,{\rm keV, exp}}^{\rm h}$\\
 & & $\rm erg~s^{-1}$ & $\rm erg~s^{-1}$ & $\rm erg~s^{-1}$ & & & $\rm erg\,s^{-1}\,cm^{-2}$ & & $\rm erg\,s^{-1}\,cm^{-2}$ \\
\hline
(A) \qsoone   & 3.229$\pm$0.003 & $78.5$ & $6.10^{+1.33}_{-1.18}$ & $1.07$ & 224$\pm$23 & 1.06$\pm$0.14 & 0.66$^{+0.14}_{-0.13}$ & 403/433 & 15.5\\
(B) \qsotwo   & 3.221$\pm$0.004 & $6.5$ & $3.54^{+1.81}_{-1.42}$ & $0.18$ & \phantom{0}67$\pm$18 & 1.73$^{+0.53}_{-0.44}$ & 0.38$^{+0.20}_{-0.15}$ & 240/263 & 2.9 \\
(C) \qsothree &  $\approx$3.170  & $-$ &  $6.64^{+1.64}_{-1.42}$ & $-$  &  136$\pm$20 & 1.11$\pm$0.17 & 0.75$^{+0.19}_{-0.15}$ & 316/301 & - \\
%\qsothree & 3.254 & -- & $\pm$ & $\pm$ \\
\hline
\end{tabular}
 \flushleft\begin{list}{}
 \item {\it Notes.} ${}^{\mathrm{a}}${ Continuum luminosity at rest-frame 2500 \ang normalised to $10^{45}$ $\rm erg~s^{-1}$. Statistical uncertainties are less than 1\%. ${}^{\mathrm{b}}$ Continuum luminosity at rest-frame 2 keV normalised to $10^{43}$ $\rm erg~s^{-1}$. ${}^{\mathrm{c}}$ Integrated \ion{C}{iv} luminosity normalised to $10^{45}$ $\rm erg~s^{-1}$. Statistical uncertainties are smaller than 1\%. ${}^{\mathrm{d}}$ \xmm/EPIC net counts in the 0.3--7 keV band. ${}^{\mathrm{e}}$ Best-fit photon index of the continuum in the baseline model (see section~\ref{x-ray analysis} for details). ${}^{\mathrm{f}}$ Observed flux at rest-frame 2 keV, corrected for Galactic absorption and normalised to $10^{-15}\,\rm erg\,s^{-1}\,cm^{-2}$.  ${}^{\mathrm{g}}$ Best-fit statistics of the baseline model: $C$ and $\nu$ represent the $C$-statistics and degrees of freedom, respectively.  \rev{${}^{\mathrm{h}}$ Expected flux (in units of $10^{-15}\,\rm erg\,s^{-1}\,cm^{-2}$) at rest-frame 2 keV obtained by assuming the $\lx-\lo$ relation published by \cite{lusso2020}.}}
 \end{list}
\end{table*}
% -----------------
\section{X-ray spectral analysis}
\label{x-ray analysis}
Before embarking on the analysis of the X-ray spectra of the quasar pair, we note that both sources are significantly fainter (by roughly an order of magnitude) than expected. In fact, as no deep X-ray observation of this field was available before the current \xmm programme, we had estimated preliminary count rates by converting the rest-frame monochromatic luminosities at 2500 \ang (Table~\ref{tab:sum}), from the WFC3 spectra, into those at 2 keV through the correlation of \citet{lr16}. Based on this, we should have collected at least several hundred net counts also from the fainter object, thus allowing a robust determination of the X-ray spectral parameters. This is clearly not the case, even considering the unfortunate position of \qsoone near a gap on the pn detector (see Figure~\ref{fig:xmmfield} and Section~\ref{obs and data red}). As a consequence, the components of the quasar pair must be either highly absorbed or intrinsically weak in the X-rays.

Given the limited data quality, we first modelled the spectra with a simple power law modified by Galactic absorption, fixed at $1.63\times10^{20}$ cm$^{-2}$ \citep{HI4PI16}. The fit is statistically acceptable for both quasars (Table~\ref{tab:sum}), and the photon index of $\gx \sim 1.7$ for \qsotwo is fully consistent with the typical values found in AGN with negligible X-ray absorption \citep[e.g.][]{2009ApJS..183...17Y,scott2011}, although the nominal 1$\sigma$ confidence range is rather large ($\sim$\,1.3--2.3). The spectral slope of \qsoone is remarkably flat instead, at $\gx = 1.06$\,$(\pm 0.14)$. The two spectra are shown in Figure~\ref{fig:xspecs} with the best-fit power-law models, and illustrate how \qsoone is detected also below 1 keV at the 3.5$\sigma$ level, ruling out the presence of a strong low-energy cut-off at $E<4$ keV in the rest frame.

We also conducted another model fit by adding an additional parameter to account for any possible column density $\nhz$ in the source frame, preserving the overall number of degrees of freedom by fixing the continuum photon index to $\gx=1.9$. The absorbed power-law model returns a nearly identical fit statistics for \qsotwo, and only an upper limit of $\nhz < 1.0 \times 10^{22}$ cm$^{-2}$ can be placed on the local column density. Conversely, for \qsoone the fit actually worsens by $\Delta C=7$, as tilting the continuum slope from the intrinsic $\gx=1.9$ to the observed $\gx\sim1.1$ requires a column density $\nhz \sim 1.6 \times 10^{23}$ cm$^{-2}$, which leaves clear positive residuals below 1 keV. A simple absorption-based model therefore fails to reproduce the X-ray spectrum of \qsoone.

Alternatively, one might suspect that the direct X-ray continuum from \qsoone is completely attenuated by a Compton-thick foreground screen, and that the observed emission is dominated by reflection from distant material. The latter interpretation is supported by a tentative line-like excess around 1.5--1.6 keV (which can be noticed in Figure~\ref{fig:xspecs} despite the coarse binning), suggesting the presence of a Fe\,K feature (6.4--7 keV in the rest frame, depending on the ionisation state). When an unresolved Gaussian profile is added to the baseline power-law model, the fit improves by $\Delta C= -11$ with the loss of two degrees of freedom. The line would have a rest energy of $E_{\rm K}=6.8^{+0.1}_{-0.5}$ keV and a rest equivalent width of $\sim$\,1 keV. This scenario, however, appears unlikely for the following reasons. First, reflection spectra from neutral matter are characterised by a much flatter continuum than observed here, if not inverted (i.e., $\gx<0$). Indeed, the spectrum of \qsoone can be accounted for by a reflection model that self-consistently includes continuum and emission lines \citep[e.g.][]{Garcia+13} only allowing for a relatively high ionisation of the gas, as also implied by the centroid energy of the putative Fe\,K feature. This poses several problems in terms of gas location and scattering geometry. Second, and even more importantly, the line-like excess is nearly coincident with one of the strongest features of the EPIC internal quiescent background, due to Al K$\alpha$ fluorescence \citep[e.g.][]{Nevalainen+05}. 

We therefore believe that the $\sim$\,1.6-keV line is an artifact associated with an imperfect background subtraction or calibration inaccuracies (or a combination of the two). Such a conclusion is corroborated by the fact that a similar feature, although with lower significance, seems to be present also in the spectrum of \qsothree\footnote{The spectrum of \qsotwo is too faint to appreciate this effect.}. We finally note that, in the absence of a direct continuum, no sensible combination of reflected and scattered \citep[e.g.][and references therein]{Gupta+21} emission can account for the observed spectral slope of \qsoone, whose origin remains unclear.
For completeness, we also analysed with the same approach the spectrum of \qsothree. The X-ray properties of this source are similar to those of \qsoone. The continuum is flat, with $\gx = 1.11\pm 0.17$, yet this time the absorbed power-law model results in a marginal statistical improvement ($\Delta C=-3$ for $\gx=1.9$), suggesting a local column of $\nhz=6^{+3}_{-2} \times 10^{22}$ cm$^{-2}$. 

% --------
\begin{figure*}
\centering\includegraphics[width=0.9\hsize]{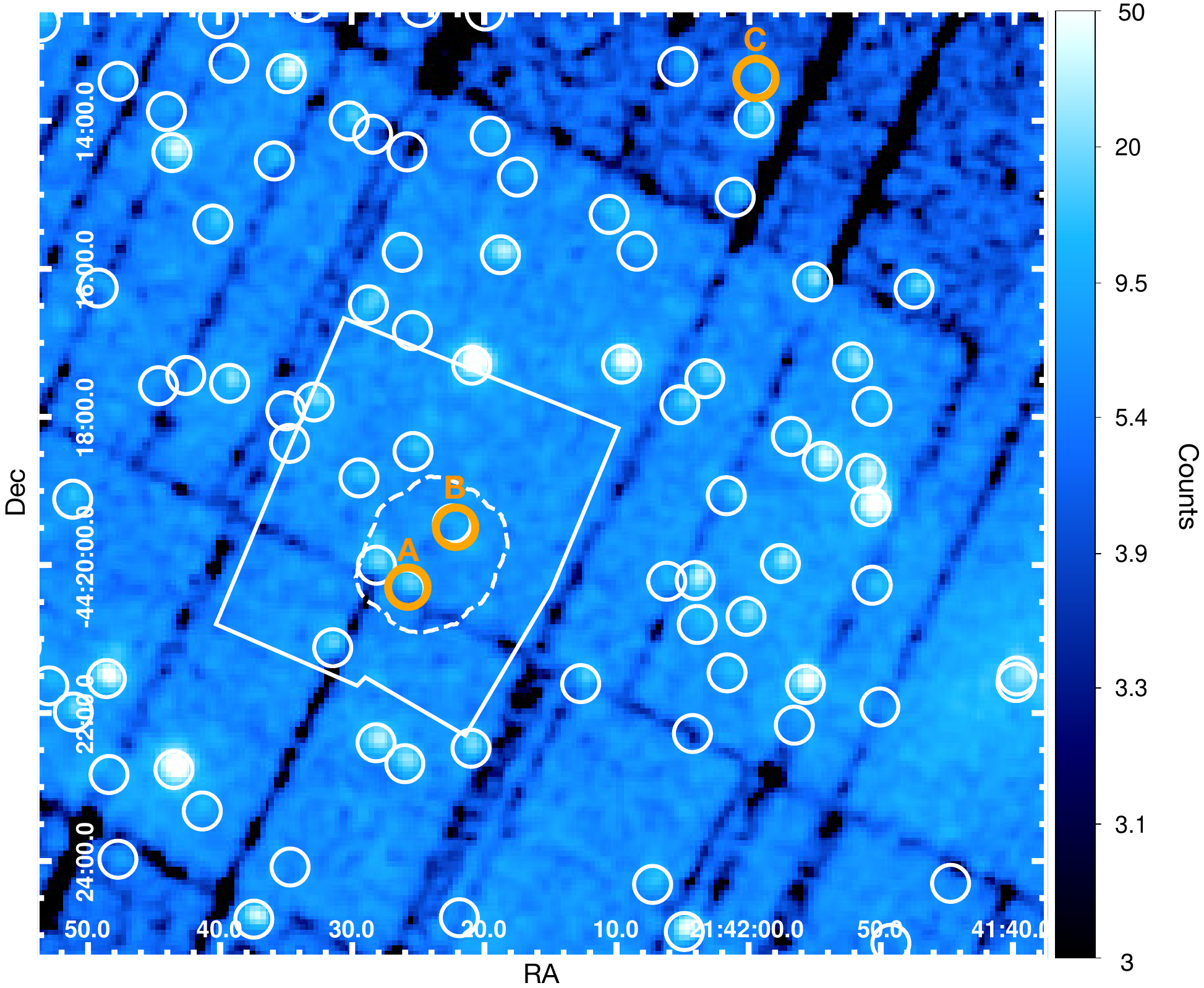}
    \caption{\rev{\xmm 0.5--7 keV band mosaic of the pn, MOS1, and MOS2 detectors. The area marked with the white solid line represents the F140W/\hst\ coverage. The white dashed lines marks the MUSE region with at least 4 hours of exposure. Orange circles represent \qsoone, \qsotwo and \qsothree. White circles mark detected sources within $10^\prime$ from the centre of the MUDF field. The image is spatially binned and smoothed through a Gaussian function. Labels A, B, and C as in Table~\ref{tab:sum}. We have considered an intensity scaling between 3 and 50 X-ray counts to balance the contrast between faint and bright sources in the field.}}
    \label{fig:xmmfield-full}
\end{figure*}
% --------
% -----------------
\begin{table*}
\centering
\caption{Summary of the properties of the seven X-ray detected sources (see section~\ref{xraysourcedet} for details) within the \hst field.} \label{tab:sum-detsources}
\begin{tabular}{lcclllcc}
\hline
\hline
\textsc{detid}$^{\mathrm{a}}$ & RA & Dec & $F_{\rm S}^{\mathrm{b}}$ & $F_{\rm H}^{\mathrm{c}}$ & \textsc{detml}$^{\mathrm{d}}$ & \hst\ ID$^{\mathrm{e}}$ & redshift \\
 & & & $\rm erg\,s^{-1}\,cm^{-2}$ & $\rm erg\,s^{-1}\,cm^{-2}$ & & & \\
\hline
108405501010054 &  325.617301   &  $-$44.333479   &   2.26 $\pm$  0.23  & 15.10 $\pm$ 1.97  & 232  & 20557  &  1.286  \\
108405501010056 &  325.636933   &  $-$44.296718   &   2.98 $\pm$  0.24  & 1.76  $\pm$ 0.97  & 204  & 2764  &    \\
108405501010083 &  325.631084   &  $-$44.352039   &   2.06 $\pm$  0.23  & 7.71  $\pm$ 1.82  & 110  & 273  &    \\
108405501010091$^\dagger$ &  325.607981   &  $-$44.338811   &   1.57 $\pm$  0.20  & 3.21  $\pm$ 1.08  & 79    & 20405  & 3.223   \\
108405501010122 &  325.622778   &  $-$44.313822   &   0.21 $\pm$  0.10  & 8.46  $\pm$ 1.60  & 43   & 2268$+$2208  &  1.249  \\
108405501010159 &  325.605758   &  $-$44.307890   &   0.58 $\pm$  0.13  & 4.79  $\pm$ 1.46  & 31   & 2604  &    \\
108405501010159 &                        &                           &                                &                              &        & 2606  &     \\
108405501010159 &                        &                           &                                &                              &        &  2622  &     \\
108405501010222$^{\dagger\dagger}$ &  325.593378   &  $-$44.324168   &   0.47 $\pm$  0.13  & 0.35  $\pm$ 0.44  & 7     & 1535  &  3.230  \\
\hline
\end{tabular}
 \flushleft\begin{list}{}
 \item {\it Notes.} $\dagger$ \qsoone. $\dagger\dagger$ \qsotwo. ${}^{\mathrm{a}}${ A unique number which identifies a detection listed in the 4XMM-DR12 catalogue. ${}^{\mathrm{b}}$ EPIC flux in the 0.5--2 keV band (in units of $10^{-15}\,\rm erg\,s^{-1}\,cm^{-2}$). ${}^{\mathrm{c}}$ EPIC flux in the 2--12 keV band (in units of $10^{-15}\,\rm erg\,s^{-1}\,cm^{-2}$). ${}^{\mathrm{d}}$ EPIC detection maximum likelihood value in the 0.2--12.0 keV band. ${}^{\mathrm{e}}$ \hst ID as reported in the online catalogue published by \citet{revalski2023}.%${}^{\mathrm{e}}$ EPIC off-axis angle: distance between the detection position and the on-axis position on the respective detector that is the smallest value of the individual camera ones.
 }
 \end{list}
\end{table*}
% -----------------
% --------
\begin{figure*}
\centering\includegraphics[width=0.9\textwidth]{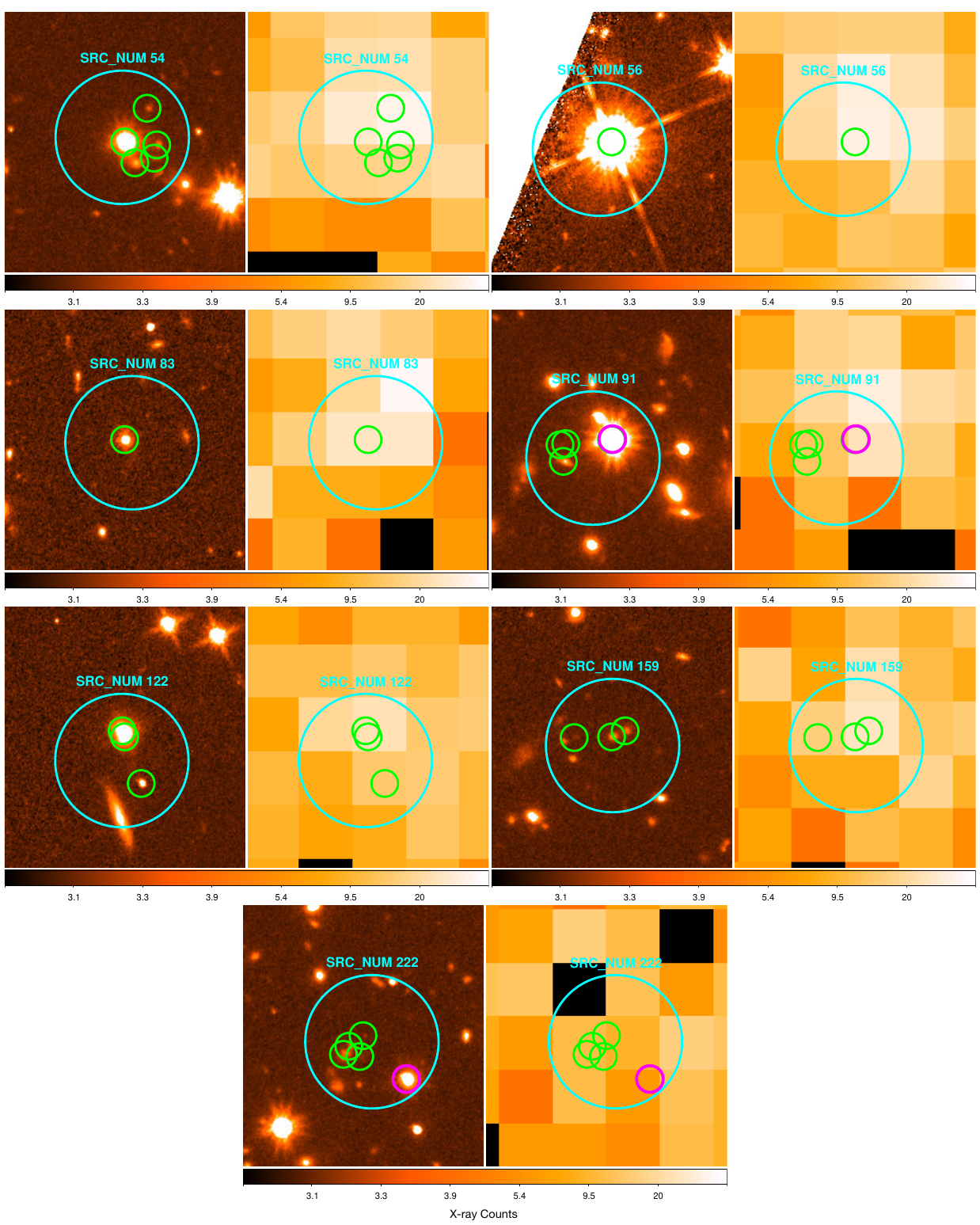}
%\centering\includegraphics[width=0.48\textwidth]{mudf-xmm-hst-54.pdf}
%\centering\includegraphics[width=0.48\textwidth]{mudf-xmm-hst-56.pdf}
%\centering\includegraphics[width=0.48\textwidth]{mudf-xmm-hst-83.pdf}
%\centering\includegraphics[width=0.48\textwidth]{mudf-xmm-hst-91.pdf}
%\centering\includegraphics[width=0.48\textwidth]{mudf-xmm-hst-122.pdf}
%\centering\includegraphics[width=0.48\textwidth]{mudf-xmm-hst-159.pdf}
%\centering\includegraphics[width=0.48\textwidth]{mudf-xmm-hst-222.pdf}
    \caption{Cutouts displaying the seven X-ray detected sources within the \hst/MUDF field. The \hst/WFC3 F140W image is on the left whilst the \xmm mosaic (pn, MOS1, and MOS2) in the 0.5--7 keV band is on the right. \rev{Each cutout has a size of $\simeq18^{\prime\prime}$ and it is centred at the \hst location of the closest counterpart based on the distance separation between the \hst and the \xmm coordinates.} In all the cutouts, the \xmm position is marked by the cyan circle ($5^{\prime\prime}$ radius, relative \xmm astrometry is $1.5^{\prime\prime}$ at $1\sigma$); the candidate counterpart(s) found within a $3^{\prime\prime}$ matching radius from the \xmm position are shown with a green, $1^{\prime\prime}$ radius, circle. The two MUDF quasars are marked by a magenta circle. The common scale bar for each cutout is a logarithmic scale of the X-ray counts, with minimum and maximum of 3 and 50, respectively. \qsoone can be safely associated with \textsc{src\_num}=91, whilst the identification of \qsotwo with \textsc{src\_num}=222 is not straightforward (see text). The foreground $z\simeq1.285$ quasar within the MUDF corresponds to \textsc{src\_num}=54, although the field is rather crowded.}
    \label{fig:xmmhstfield-cutouts}
\end{figure*}
% --------

\subsection{Source detection within and beyond the MUDF field}
\label{xraysourcedet}
Given the presence of nearby point-like or extended (i.e. more than the \xmm point spread function)  emission structures in the MUDF, we searched for additional X-ray detections within the F140W/\hst\ field of view. As the \xmm field is far larger than the F140W/\hst\ coverage, we extended this search to a distance of 10$^{\prime}$ to provide a detection list that roughly matches the distance of the quasar \qsothree from the MUDF field.
We considered the centre of the MUDF/\hst field at the coordinates $(325.6, -44.33)$ degrees and we cross-matched them with the 4XMM--DR12 source catalogue \citep{webb2020}. 4XMM--DR12 is the fourth generation catalogue of serendipitous X-ray sources available online and contains 939,270 X-ray source detections (630,347 unique X-ray sources) made public on or before 2021 December 31\footnote{\url{http://xmmssc.irap.omp.eu/Catalogue/4XMM-DR12/4XMM_DR12.html}}. The net sky area covered when accounting for overlaps between observations is $\sim$1283 deg$^2$, for a net exposure time $\geq$1 ks. To define a reasonably clean sample, we have applied the following quality cuts from the 4XMM--DR12 catalogue: \textsc{sum\_flag}$<$3 (low level of spurious detections), and \textsc{confused}$=$0 (low probability of being associated with two or more distinct sources)\footnote{For more details, the interested reader should refer to the 4XMM catalogue user guide at the following website \url{http://xmmssc.irap.omp.eu/Catalogue/4XMM-DR12/4XMM-DR12_Catalogue_User_Guide.html}.}.
\rev{For the cross-match between the \hst/MUDF and the 4XMM catalogues, we considered the corrected RA and Dec listed in the 4XMM catalogue after the application of a statistical correlation of the \textsc{emldetect} coordinates with the USNO B1.0, 2MASS or SDSS (DR8) optical/IR source catalogues using the \sas task \textsc{catcorr} (i.e. field rectification)\footnote{\url{http://xmmssc.irap.omp.eu/3XMM-DR4/UserGuide_xmmcat.html\#Astrom}}. Therefore, the centroid is not necessarily the same as the one defined by the X-ray peak flux. This is also consistent with the  source extraction performed by hand, as we considered the optical coordinates since the X-ray data are such that we cannot clearly identify the X-ray source position.}
The above search led to a sample of 119 X-ray detections, about 80\% of which are flagged as ``good'' (94 detections with \textsc{sum\_flag}=0, i.e. no warnings raised in any of the EPIC bands), whilst the remaining 25 detections have at least one warning flag raised, although the detection is considered reliable.

Inside the MUDF/\hst field we found seven \rev{detections} (see Figure~\ref{fig:xmmfield-full}), three of which fall within the narrower MUSE field: \qsoone and \qsotwo (although the identification with the latter is more controversial, see below), and a lower redshift quasar at $z\simeq 1.285$ (cyan cross in Figure 8 by \citealt{fossati2019}), which lies close to the edge of the MUSE FoV in close spatial proximity to a $z\simeq0.67$ galaxy group (see Section 6.2 in \citealt{fossati2019} for more details). The low number of detected sources within the field is expected given the poorer \xmm angular resolution (6 arcsec FWHM, or equivalently 15 arcsec half-energy width) as compared to \hst.  

Table~\ref{tab:sum-detsources} summarises the properties of the seven X-ray detected sources within the \hst field: the unique number (\textsc{detid}), which identifies a detection as listed in the 4XMM--DR12 catalogue (column 1); RA and Dec relative to the X-ray position (columns 2 and 3); the observed flux with uncertainties in the 0.5--2 keV band (column 4), computed as the sum of the EPIC 0.5--1 and 1--2 keV energy bands (flagged in the catalogue as band 2 and 3, respectively); the observed flux with uncertainties in the 2--12 keV band (column 5), computed as the sum of the EPIC 2--4.5 and 4.5--12 keV energy bands (flagged in the catalogue as band 4 and 5, respectively); the EPIC detection maximum likelihood (\textsc{ep\_8\_det\_ml}) value in the 0.2--12.0 keV band (column 6). We also included the \hst\ catalogue IDs (column 7) for objects covered by the F140W and the redshift (column 8).

Of the seven X-ray detections within the \hst field, only two have a single \hst counterpart within $3^{\prime\prime}$ from the \xmm location, i.e. \textsc{src\_num}\footnote{The \textsc{src\_num} represents the (decimal) source number in the individual source list for the relative \xmm observation (i.e. the last digits of \textsc{detid}).}=56 and 83 (see Figure~\ref{fig:xmmhstfield-cutouts}). We do not have any spectral coverage for these two objects. \textsc{src\_num}=83 shows slightly extended emission as well as faint, possibly offset emission in F336W, so it could be either a star or another AGN, but the correct classification requires additional spectroscopy. 
The object with \textsc{src\_num}=56, located at the extreme edge of the F140W coverage, is a foreground Milky Way star (probably a Wolf-Rayet).
Counterparts for the X-ray detections flagged with \textsc{src\_num}=54, 91, 222 and 122 are, with different degrees of confidence, the $z\simeq1.286$ quasar in close spatial proximity to a $z\simeq0.67$ galaxy group, \qsoone, \qsotwo, and another $z\simeq1.249$ quasar, respectively, although the emission of the latter is closely blended with another \hst\ object (\hst ID=2208). In particular, the association of \textsc{src\_num}=222 with \qsotwo is not straightforward, as several other optically detected sources lie closer to the coordinates of the X-ray detection. We recall that the spectral extraction region of $12^{\prime\prime}$ radius definitely encompasses any X-ray emission from \qsotwo, but it possibly includes also some contribution from these nearby objects. We therefore conclude that the X-ray fluxes we have derived for \qsotwo from the spectral analysis, despite being already anomalously low, should be likely treated as upper limits. Depending on the actual level of contamination to its spectrum (Fig. \ref{fig:spectra}), we cannot even exclude that \qsotwo is heavily obscured in the X-rays. Sub-arcsecond accurate X-ray positions would be required to clarify this issue but, to date, no \chandra observations of the MUDF exist. Also for the remaining detection (\textsc{src\_num}=159), there is ambiguity on the most probable counterpart. 

% NOTE on the flux differences between catalogue and spectral analysis for qso1 and qso2
As a consistency check, we compared the flux values obtained for \qsoone\ and \qsotwo from the spectral analysis with the catalogued ones. 
%From the spectral analysis of \qsoone, we obtain $2.1\times10^{-15}$ erg s$^{-1}$ cm$^{-2}$ and $6.1\times10^{-15}$ erg s$^{-1}$ cm$^{-2}$ in the soft (0.3--2 keV) and hard (2--7 keV) bands with statistical 1$\sigma$ uncertainties of $\sim$12 and 15 per cent, respectively; whilst we have a flux of $0.8\times10^{-15}$ erg s$^{-1}$ cm$^{-2}$ in both bands for \qsotwo, with statistical 1$\sigma$ uncertainties of 25 and 45 per cent, respectively.  
Based on the best-fit power-law model of \qsoone (Table \ref{tab:sum}), we obtain a flux of $8.2\times10^{-15}$ erg s$^{-1}$ cm$^{-2}$ with statistical 1$\sigma$ uncertainty of \rev{18}\% over the 0.3--7 keV band, whilst we have $1.6\times10^{-15}$ erg s$^{-1}$ cm$^{-2}$ with 1$\sigma$ uncertainty of $\sim$\rev{30}\% for \qsotwo. These spectral fluxes are larger by factors of 1.5 (\qsoone) and 1.1 (\qsotwo) than the catalogued values relative to the full EPIC band (0.2--12 keV, band 8). \rev{Given the large uncertainties, for \qsotwo the agreement remains good (i.e., within 1$\sigma$) even after extrapolating the flux computed through our spectral analysis to the 0.2--12 keV EPIC band. For \qsoone, instead, there is a formal inconsistency at the 2.2$\sigma$ level with respect to the 4XMM catalogue, almost entirely arising above 2 keV.}  
%The agreement is generally good (i.e., within the uncertainties) in the soft band for both sources, but above 2 keV the fluxes computed through our spectral analysis are brighter by a factor of $\approx$2 with respect to the ones listed in the 4XMM catalogue. 
There are several possible explanations to alleviate this apparent discrepancy. Although we performed the spectral fits over the 0.3--7 keV energy range, almost no counts are detected beyond 5 keV for both quasars. Therefore, extrapolating the best-fit power law at higher energies (especially for a very flat $\Gamma$ as found for \qsoone) \rev{clearly leads} to an overestimate of the hard-band flux. On the other hand, the catalogued \rev{band-8} fluxes for all the X-ray detected objects assume a fixed power-law source spectrum \citep[$\Gamma_{\rm X}=1.42$;][]{webb2020} %in each narrow energy band 
irrespective of the \rev{actual} broadband spectral shape. The EPIC total-band flux is then computed as the %cumulative flux from five independent narrow bands. 
\rev{weighted average between the three detectors.} 
This procedure, however, can introduce some systematic uncertainty at low fluxes. Indeed, the band-8 flux of \qsoone derived from the pn image is more than two times larger than those obtained from the MOS ones. This difference is entirely due to band 5 (4.5--12 keV), where no source counts are detected by either MOS camera, for a nominal zero flux. This brings down the combined EPIC band-5 (hence band-8) flux of \qsoone. Incidentally, the catalogued band-8 pn flux for this quasar is $9.2\times10^{-15}$ erg s$^{-1}$ cm$^{-2}$, which is broadly consistent with our spectral estimate. 

We finally note that, for a simple pivot effect, the above hard-band-related systematics have little impact on the determination of the monochromatic flux at rest-frame 2 keV. \rev{In fact, the fit anchors the continuum power law to the soft band, where most of the counts are found, so that the 2-keV flux is barely sensitive to the actual spectral slope.} If anything, if we were to adopt the broadband 4XMM fluxes as reference, both quasars would be even fainter in the X-rays than assumed in the following discussion, so our main results are completely unaffected.

\section{Ultraviolet spectral analysis}
\label{Ultraviolet spectral analysis}
The MUSE and \hst spectra are fitted by using the publicly available package for spectral fitting \qsfit\ \citep{calderone2017}.
The observed emission lines in quasar spectra are reproduced by a broad (FWHM\,$>$\,2,000 \kms) profile and, when required, an additional narrow component (FWHM\,$<$\,2,000 \kms) is included, whilst the continuum considers contributions from both the iron UV complex and the nuclear ionising continuum (i.e. disc emission). To improve the residuals, we also
considered a set of ``unknown'' emission lines, i.e. emission not associated with any known line (see Section 2.7 in \citealt{calderone2017}). These components account for the lack of an iron template in the wavelength range 3100–-3500 \AA, or for possible asymmetric profiles in known emission lines. The spectra were corrected for Galactic extinction by using the $\ebv$ value of 0.017 from \citet{sf2011} and the parametrisation by \citet{CCM1989} and \citet{odonnel1994}, with a total to selective extinction parameter $R_V$\,$=$\,3.1 \citep{calderone2017}.

\revs{At rest-frame wavelengths bluer than $\simeq1216$ \AA,} absorption from intergalactic \ion{H}{i} attenuates the quasar flux, both in the Lyman series, and in the Lyman continuum \citep[e.g.][]{2009ApJ...705L.113P}. \revs{A correction for IGM absorption is thus required to properly retrieve both the \ion{Ly}{$\alpha$} and the continuum emission \citep[e.g.][]{lusso2015}. We also have a gap in the range 1362--1421 \AA\ due to the blocking filter that avoids the light of the sodium laser of the adaptive optics system.}
Therefore, to perform the spectral fit, we conservatively neglected all the wavelengths shorter than 1450 \ang at rest in the MUSE data.

\begin{figure}
	\includegraphics[width=\hsize]{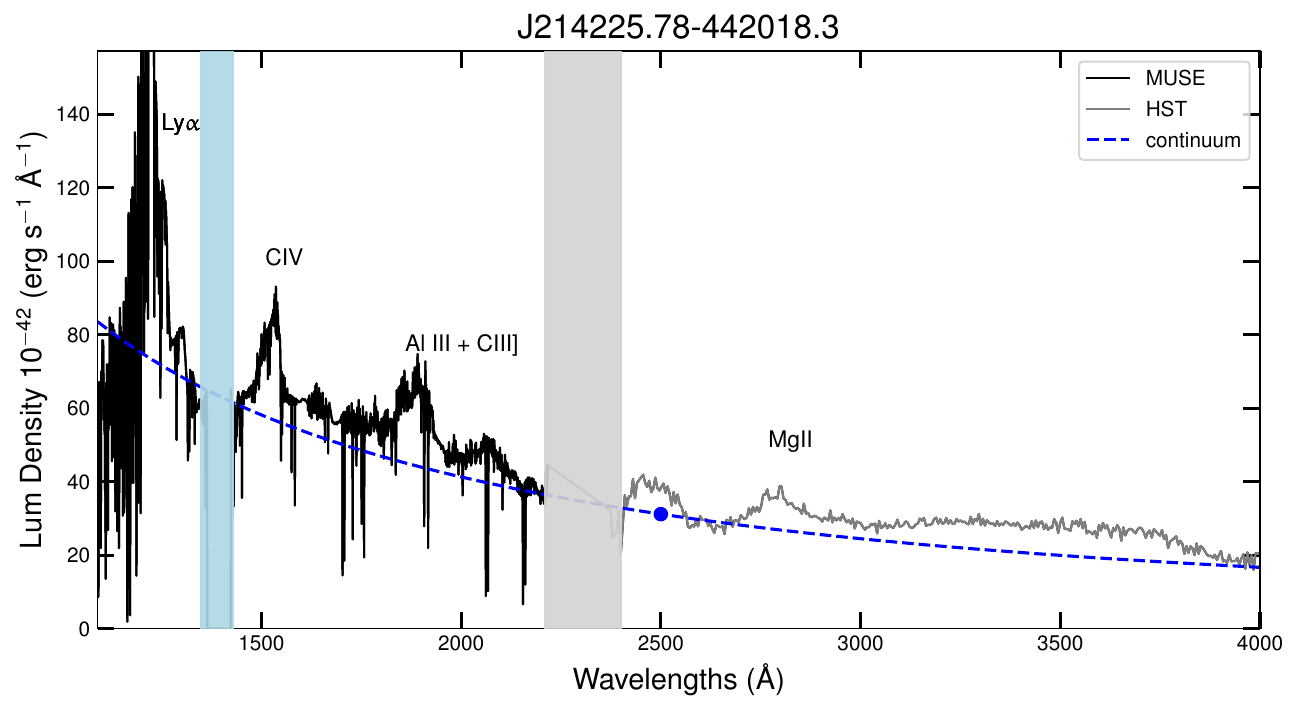}
	\includegraphics[width=\hsize]{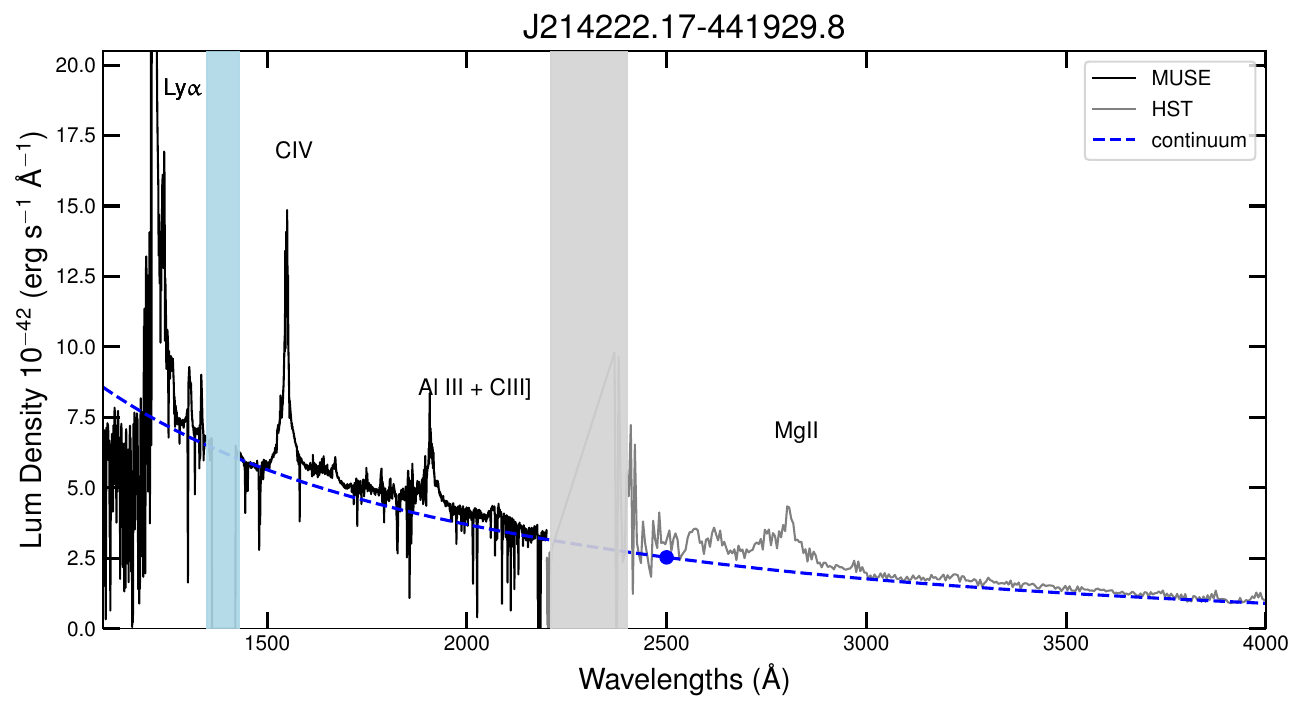}
    \caption{MUSE and \hst spectra for the two quasars in the MUDF shown at rest wavelengths. The best-fit nuclear continuum is shown with the blue dashed line. All the other components considered in the spectral fit (e.g. iron complex) are not shown in this figure for visual clarity. The blue point marks the rest-frame continuum luminosity at 2500 \ang. \rev{Grey shaded regions mask the gap between the MUSE and \hst\ spectra}, whilst the cyan shaded region masks the portion of the spectrum with missing data due to the blocking filter that avoids the light of the sodium laser of the adaptive optics system. The main emission lines are labelled.}
    \label{fig:spectra}
\end{figure}

The first few hundred angstroms of the \hst spectrum are also neglected, as the extreme blue edge of the detector has some artifacts that produce erratic changes at the shortest wavelengths (i.e. around 2500 \ang, see Figure~\ref{fig:spectra}). 

% flux calibration
Regarding the flux calibration, the MUSE detector has been extremely well calibrated over the last several years. Data have an absolute flux calibration accuracy of 5--10\% per single exposure, leading to uncertainties of a few per cent (dominated by systematics) on the final co-addition of several hundred exposures. 
Therefore, we adopted the MUSE data as a reference to match the \hst spectra. \rev{On average}, the \hst and MUSE data are both well matched for all the sources observed in the MUDF, with flux differences within $\sim10\%$. Given the very few emission-line free windows in the \hst data, we prefer to adopt the same slope of the MUSE spectra for the \hst data, although the slopes agree within uncertainties when fitted separately. Moreover, %by fitting the MUSE and \hst data separately, 
the MUSE and \hst continuum level was the same in the case of the brighter quasar, \rev{whilst an offset was applied to the \hst data of \qsotwo\ to better match the underlying MUSE continuum. This offset is nonetheless very small, less than a factor of 1.2 in flux}.
Figure~\ref{fig:spectra} presents the MUSE and \hst spectra of the two quasars in the MUDF. Missing data in the MUSE spectrum are marked with a cyan shaded region. Grey shaded regions mask \rev{the gap between the MUSE and the \hst\ spectra}. The best-fit nuclear continuum is shown with the blue dashed line ($\alpha_\nu=-0.67\pm0.05$ and $\alpha_\nu=-0.57\pm0.05$ for  \qsoone\ and \qsotwo, respectively), whilst the blue point refers to the rest-frame continuum luminosity at 2500 \ang.  

% --------
% --------

\section{Results and Discussion}
\label{results and discussion}
\begin{figure}
	\includegraphics[width=\hsize]{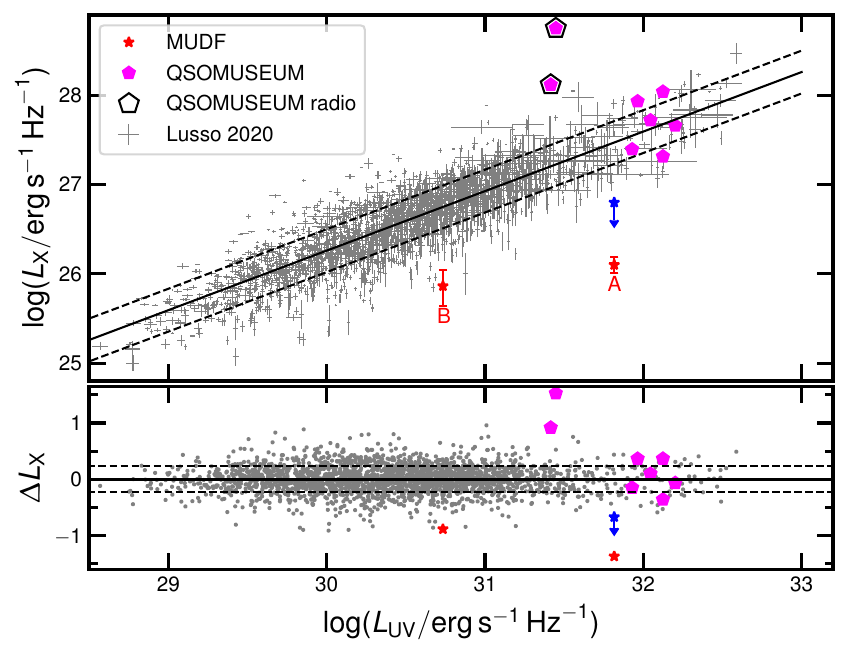}
    \caption{Rest-frame monochromatic luminosity at 2 keV ($\lx$) against the one at 2500 \AA\ ($\lo$) for the two quasars in the MUDF (red stars). The blue arrow represents the X-ray upper limit for \qsoone\ when an intrinsic column density is assumed (see Section \ref{x-ray analysis} for details). The grey symbols represent the sample of $\sim$\,2,400 quasars from \citet{lusso2020}, with the relative best-fit regression (black solid line). The dashed lines trace the 1$\sigma$ dispersion, 0.24 dex. Magenta pentagons represent $z\simeq3$ quasars from the QSO MUSEUM survey \citep[][]{arrigoni2019a} with X-ray data available from the archive. Labels A and B as in Table~\ref{tab:sum}.}
    \label{fig:lolx}
\end{figure}

% lo-lx
Figure~\ref{fig:lolx} shows the rest-frame monochromatic luminosity $\lx$ against $\lo$ for the two quasars in the MUDF. The blue arrow is the X-ray flux for \qsoone\ when an intrinsic column density and a fixed $\gx=1.9$ are assumed. We recall that the latter spectral fit is statistically worse than the case where no intrinsic absorption is allowed for (see Section \ref{x-ray analysis} for details), yet we adopt the resulting X-ray flux as a conservative upper limit. Likewise, the flux of \qsotwo should be likely considered as an upper limit due to the possible contamination from nearby objects (Section \ref{xraysourcedet}). Figure~\ref{fig:lolx} also shows the sample of quasars from \citet{lusso2020}, with the relative best-fit regression line \rev{(with a slope $\gamma$\,$=$\,0.667\,$\pm$\,0.007 and intercept $\beta$\,$=$\,6.25\,$\pm$\,0.23) for comparison. The dashed lines trace the 1\,$\sigma$ dispersion, 0.24 dex}. 
% description of the sample
The \citet{lusso2020} sample is composed by 2,421 optically-selected quasars (the majority from SDSS) with X-ray data from \xmm\ and \chandra, and it spans a redshift range 0.01\,$\leq$\,$z$\,$\leq$\,7.54, with a mean (median) redshift of 1.4 (1.3). These sources were selected to represent typical quasars, thus to have minimal host-galaxy contamination (especially important for the $z$\,$<$\,0.7 AGN) and minimal gas and dust absorption. The Eddington bias is also taken into account. Details about the sample selection are provided in their Section~5. 

From an observational perspective, the $\lx-\lo$ relationship provides a robust estimate of the quasar expected X-ray luminosity for any given UV luminosity, thus determining the range of soft X-ray emission for typical (i.e., non-broad absorption line, non-jetted, with minimal dust and gas absorption) quasars or, vice-versa, pinpointing peculiar objects (e.g. X-ray weak, with strong radio jets, or extremely red). Physically, \rev{this relation} indicates a strong link between the accretion disc (emitting in the UV) and the X-ray corona \citep[e.g.][]{avnitananbaum79,zamorani81,vignali03,steffen06,just07,lusso2010,martocchia2017}. This non-linear correlation is tight \citep[$\leq$\,0.24 dex of scatter;][]{lr16}, with a slope independent of redshift, suggesting that the connection between the disc and the corona must exist in AGN across cosmic time \citep[e.g.][]{nicastro2000,merloni2003,lr17,arcodia2019}.

% --------
% --------

Both \qsoone\ and \qsotwo\ deviate from the $\lx-\lo$ relation, with faint X-rays with respect to quasars at similar UV luminosities. This is even more striking for \qsoone, whose X-ray emission lies at $>$\,3$\sigma$ below the $\lx-\lo$ relation. This source shows an X-ray bahaviour similar to the X-ray weak quasars at $z=3.0-3.3$ published by \citet{nardini2019}, who discovered that $\approx$\,25\% of the quasars in their sample present an X-ray emission much weaker than expected, by factors of 3 or higher. Such an X-ray weak fraction is significantly larger than those previously reported for radio-quiet, non-broad absorption line (BAL) quasars at lower redshift and luminosity ($\approx$\,8\%, e.g. \citealt{Gibson2008,pu2020}, see also \citealt{timlin2020}). Their X-ray weak quasars display a flat photon index ($\gx<1.6$) with no clear evidence of X-ray absorption in the \xmm\ spectra. \rev{The expected fluxes at rest-frame 2 keV, if the MUDF quasars were to follow the $\lx-\lo$ relationship published by \cite{lusso2020}, are listed in Table~\ref{tab:sum}. The observed X-ray fluxes at 2 keV are fainter by a factor of $\sim$23.5 and 7.6 than the expectations for \qsoone\ and \qsotwo, respectively.}

% ------- QSO MUSEUM
We also included in Figure~\ref{fig:lolx} all quasars  within the QSO MUSEUM survey \citep{arrigoni2019a} with an X-ray observation available. 
The QSO MUSEUM (Quasar Snapshot Observations with MUse: Search for Extended Ultraviolet eMission) sample consists of 61 quasars at $3.03 < z < 3.46$ (median redshift $z=3.17$) with absolute $i-$band  magnitude normalized at $z = 2$ \citep{ross2013} in the range $-29.7\leq M_i(z=2)\leq -27.0$ and different strengths of radio emission. All these quasars have been observed with MUSE to characterise the physical properties of the CGM and IGM in emission (chiefly through the \lya\ transition) associated to these bright quasars.

We have searched for any X-ray observations (either pointed or serendipitous) of all the QSO MUSEUM sources in the \xmm and \chandra archives. Besides \qsoone and \qsothree, which are also part of the sample, we found other eight objects with publicly available X-ray data. Two sources belong to the \citet{nardini2019} sample, and are not re-analysed here. For the remaining six quasars, we retrieved and reduced the archival data with the standard procedures, and analysed the extracted spectra as described in Section \ref{x-ray analysis}. More details on the targets and on the observations are provided in Appendix \ref{app.A}. 
% L2500 
Their rest-frame luminosities at 2500 \ang are computed from the extrapolation of the continuum best fit of the MUSE spectra with a similar methodology as discussed in Section~\ref{Ultraviolet spectral analysis} (no \hst data are available for this sample). 

% lx-lo
The majority of quasars lie in agreement with the $\lx-\lo$ relation. ID49 and ID54 are bright radio sources, therefore their X-ray emission includes an additional synchrotron component due to the radio jet. Only \qsoone (i.e. ID22) and \qsotwo appear to be X-ray weak, leading to an X-ray weak fraction of 20\%, which raises to 25\% if we exclude the radio-bright quasars. Given the current data, we thus conclude that the fact that both quasars in the MUDF are X-ray weak is likely to be serendipitous and mostly driven by the higher probability to observe an X-ray weak quasar at these redshifts. 

\begin{figure}
	\includegraphics[width=\hsize]{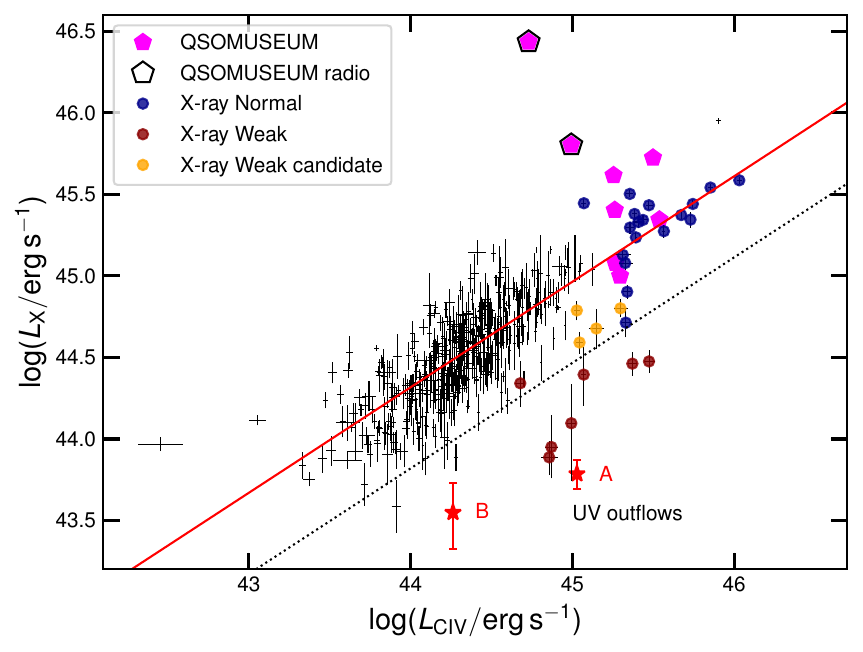}
    \caption{X-ray luminosity at rest-frame $2$ keV as a function of the total integrated \ion{C}{iv} line luminosity. Black symbols describe a sample of quasars with SDSS and XMM data at lower redshift and luminosity, with respect to the MUDF quasars, with a robust measurement of the integrated \ion{C}{iv} line emission (Signorini et al., in preparation). The red solid line is the best-fit regression obtained with the SDSS-XMM sample and the X-ray Normal $z\simeq3$ quasars. The dotted line is 3$\times$ the intrinsic dispersion on the best-fit relation, $\simeq$\,0.5 dex. Blue, brown and orange symbols represent X-ray normal, weak and weak candidates, respectively, following the definition in Section 2.3 in \citet{lusso2021}. Red stars are \qsoone\ and \qsotwo. Magenta pentagons represent $z\simeq3$ quasars from the QSO MUSEUM survey \citep[][]{arrigoni2019a} with X-ray data available from the archive. Labels A and B as in Table~\ref{tab:sum}.}
    \label{fig:lcivlx}
\end{figure}

% lx-lciv
\rev{The MUDF quasars are located in the lower right corner of the $\lx-L_{\rm C\,IV}$ plane, as shown in Figure~\ref{fig:lcivlx}, consistently to what is observed by \citet{nardini2019} and \citet{lusso2021} in their X-ray weak quasar sample. By analysing the properties of the high-ionisation \ion{C}{iv}\,$\lambda$1549 broad emission line in connection with the X-ray emission, \citet{lusso2021} observed a tight correlation (with a slope of 0.647$\pm$0.001 and an observed dispersion of $\simeq0.2$ dex), between the X-ray luminosity at rest-frame $2$ keV and the total integrated \ion{C}{iv} line luminosity (see their Figure~14), in a sample of $\simeq$\,1,800 quasars. Their sample was selected to fulfil all the quality criteria discussed by \citet[][]{lusso2020}, thus non-broad absorption line, non-jetted, with minimal deviation due to absorption at both UV and X-rays, so it is representative of typical blue quasars in the redshift range $1.7\lesssim z\lesssim 3.3$. The observed $\lx-L_{\rm C\,IV}$ relation implies a strong link between the relative strength of the X-rays with respect to both the UV continuum and the \ion{C}{iv} line emission (see their Figure 15, their Section~4.5 and relative discussion for more details).
%Bright quasars that appear weaker in the X-rays with respect to the expected emission (if they were to follow the $\lx-\lo$ relation), present a brighter \ion{C}{iv} line flux compared to sources at similar X-ray luminosities (considering the $\lx-L_{\rm C\,IV}$ relation as a reference). In other words, even though the \ion{C}{iv} line emission looks quite ordinary, it should have been much fainter for X-ray weak quasars if they were to follow the $\lx-L_{\rm C\,IV}$ correlation.
\citet[][see also \citealt{nardini2019,lusso2021}]{trefoloni2023} argue that X-ray weakness could also be interpreted in a starved X-ray corona picture, connected with an ongoing wind phase that may extend to kpc scales (e.g. \citealt{bischetti2017,vietri2018,zappacosta2020,temple2023}, and references therein).
If the wind is ejected in the vicinity of the black hole, the UV light that reaches the corona will be depleted, depriving the corona of \rev{seed} photons and eventually generating an X-ray weak quasar. Yet, in the quasar luminosity regime, there will still be an ample reservoir of ionising photons that produce the \ion{C}{iv} emission observed in the X-ray weak quasars with respect to typical sources of similar X-ray luminosities (see Section~5 in \citealt{lusso2021}).  }

% --------
% --------

\rev{We also note that the MUDF is not covered by any radio facility, at present. We searched in the FIRST, NVSS, ALMA, NRAO (EVLA, GBT, GMVA, VLA, VLBA), and LOFAR archives finding no matches, so we do not have any information regarding whether MUDF quasars are radio bright or not. Radio data may help in the interpretation of the X-ray data, since synchrotron emission may emit at X-ray energies as well \citep[see e.g.][]{page2005,zhu2020}. Yet, since both MUDF quasars are X-ray weak, we may guess that they are also radio quiet, even accounting for possible X-ray variability. While possible X-ray obscuration of the coronal emission can occur at circumnuclear scales (broad line region and/or torus), a Compton-thick column at host-galaxy scales would be required to fully absorb the emission of an extended jet, which seems highly unlikely.}

Summarising, we found that both MUDF quasars are intrinsically X-ray underluminous with respect to typical quasars at similar UV and \ion{C}{iv} luminosities and redshifts. \rev{We do not know whether extreme X-ray variability (e.g. \citealt{timlin20b}) may play a role, as multiple X-ray observations are not available, but this is very unlikely to occur {\it in phase} for both sources. We cannot thus exclude that the observed, simultaneous X-ray deficit could have instead a causal, common physical origin for both MUDF quasars. This might be related to the SMBH growth history (i.e. duty cycle), the environment, and/or the small-scale accretion physics of the two quasars, but we cannot investigate these scenarios any further at this stage, until more data are collected (e.g. deeper X-ray and radio observations). Nonetheless, we found consistent results with previous works in the literature that observed a high fraction ($\sim25$\%) of X-ray weak quasars at $z\simeq 3$ \citep[e.g.][]{nardini2019,zappacosta2020,lusso2021} with respect to lower redshift AGN samples ($\simeq8$\%, see e.g. \citealt{Gibson2008,pu2020,timlin2020}). These works point towards a high incidence of outflowing gas in the X-ray weak population with respect to the X-ray normal one at high redshifts. Alternatively to starving the corona, such outflows could also provide an additional source of X-ray obscuration (e.g. \citealt{huang23}), although absorption does not clearly emerge from our spectral analysis.} %consistent with what is observed in the UVES/ESO spectra of the MUDF quasars, which display associated absorption features again characteristic of the presence of outflowing material at large scales. }

% --------
\begin{figure*}
	\includegraphics[width=0.48\textwidth]{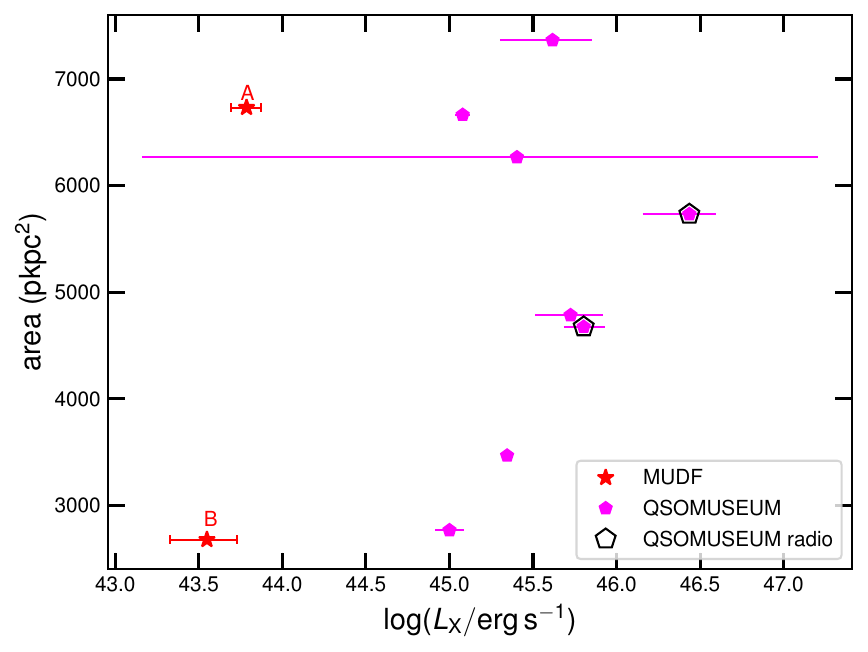}
	\includegraphics[width=0.48\textwidth]{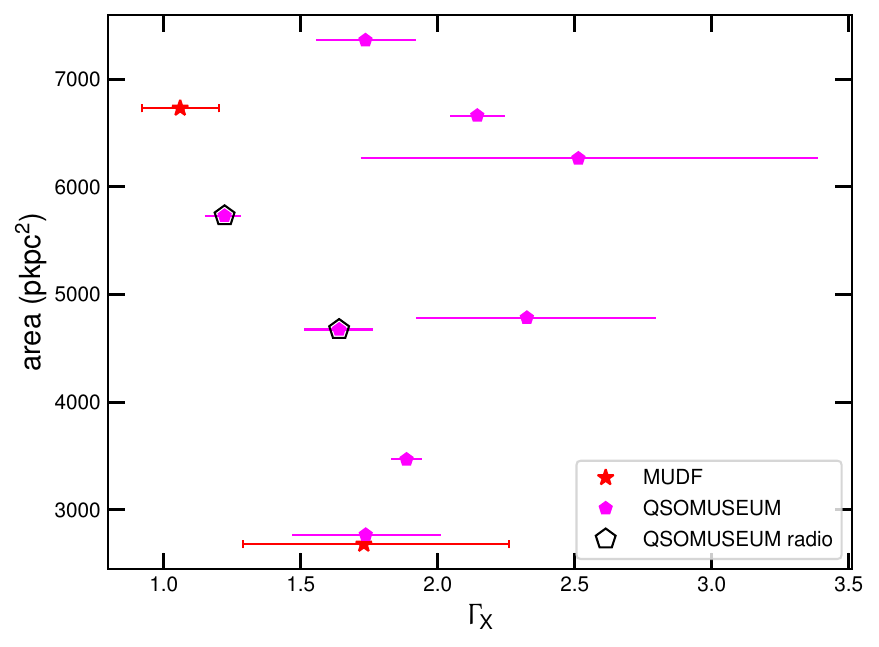}\\
 	\includegraphics[width=0.48\textwidth]{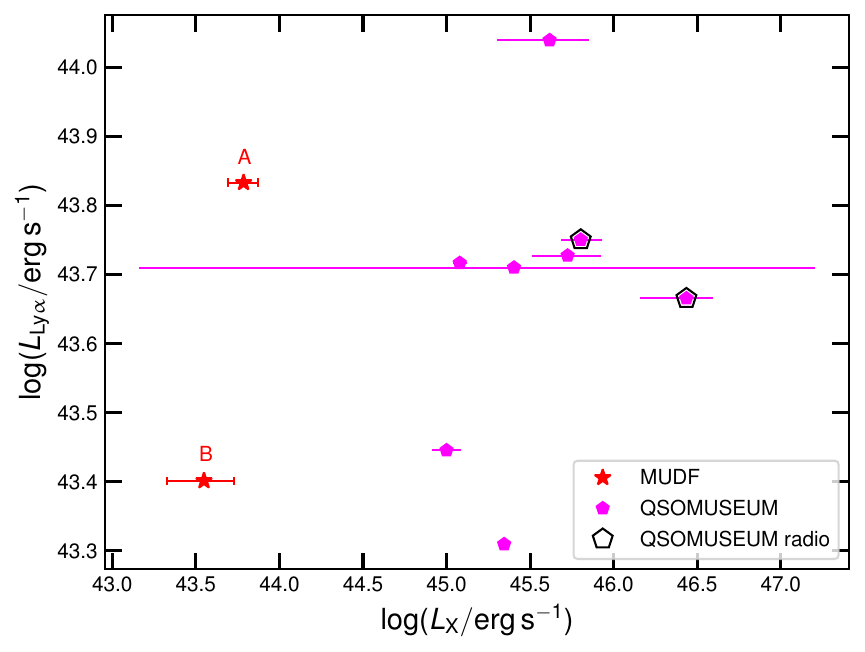}
    \caption{Upper left panel: area of the extended \lya\ emission as a function of the quasar luminosity at rest-frame 2 keV. Key symbols as in Figure~\ref{fig:lolx}. The large error bars on $L_{\rm X}$ for ID27 take into account the large uncertainties on the photon index observed for that source (see Appendix~\ref{app.A}). \rev{Upper right panel: area of the extended \lya\ emission as a function of the X-ray photon index}. Bottom panel: \lya\ integrated emission as a function of the luminosity at rest-frame 2 keV. We note that the MUDF and QSO MUSEUM \lya\ values have been computed under slightly different assumptions. Yet, the lack of a correlation holds even after accounting for minor systematics.}
    \label{fig:areax}
\end{figure*}
% --------
% --------
\begin{figure*}
	\includegraphics[width=0.48\textwidth]{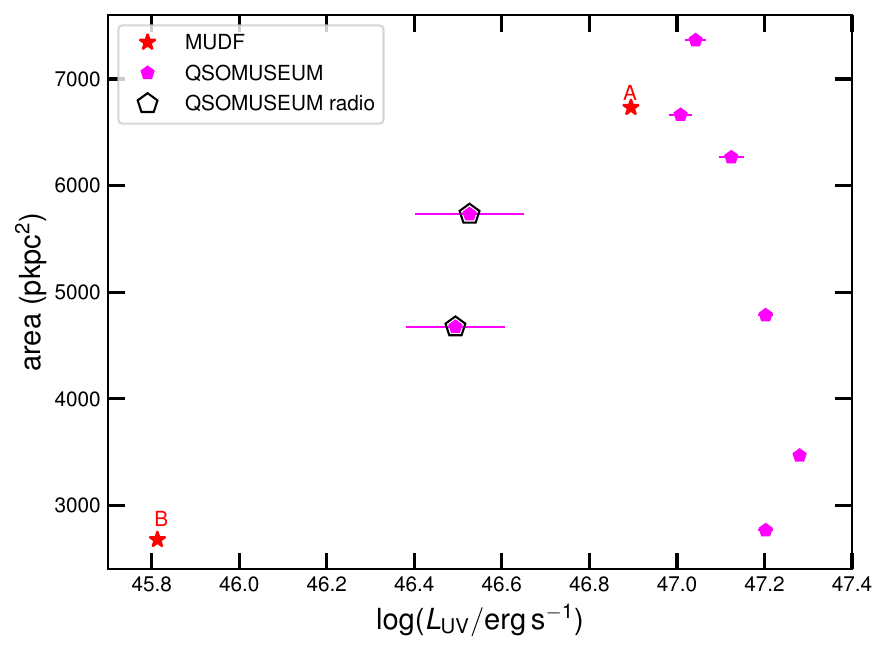}
 	\includegraphics[width=0.48\textwidth]{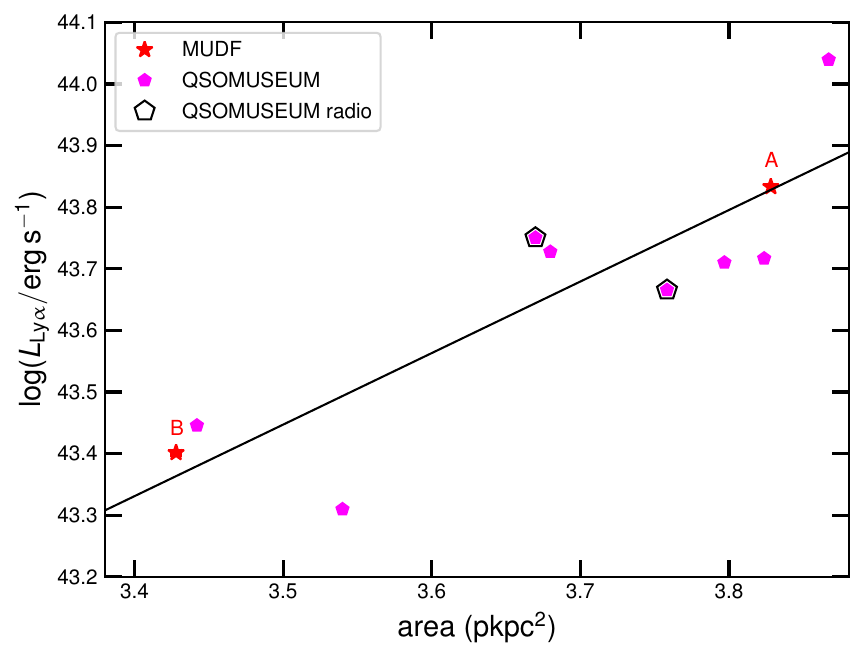}
    \caption{Left panel: area of the extended \lya\ emission as a function of the quasar luminosity at rest-frame 2500 \AA. \rev{Right panel: \lya\ integrated emission as a function of the area. The MUDF and QSO MUSEUM \lya\ values have been computed under slightly different assumptions, but any relative shifts are within the scatter of the relation (see section~\ref{area-lum discussion} for details). The black solid line is the best regression fit of the sample.}}
    \label{fig:area}
\end{figure*}
% --------

% \\\textbf{NOTE1: this plot can include ALL QSO MUSEUM quasars, not the X-ray detected solely (Fabrizio let me know what you think about it). NOTE2: the top axis is not correct, I need the right equation used by Arrigoni-Battaia in his paper to derive $Mi(z=2)$}
\subsection{The nebula-quasar connection}
\label{area-lum discussion}
\cite{mackenzie2021} found a statistically significant correlation between the surface brightness (SB) of the nebula observed in \lya\ emission at the redshift of quasars and the luminosity of the quasar in both UV continuum and \lya. Specifically, they observed that fainter quasars in the UV have on average smaller, less-luminous nebulae, with lower outer SBs (see their Figure 6). They defined the size of a nebula as the largest projected distance (maximum extent) between pixels within the 3D mask of the optimally extracted MUSE images. This mask comprises all the voxels (volume pixels) identified as part of the \lya\ nebulae, which is then used to
define the size and properties of the nebulae (see their Section 3.1 for more details). This definition represents a good metric of the morphology of the nebulae only if the asymmetry in the extended \lya\ emission is limited, which is often not the case \citep[see e.g.][]{borisova2016,arrigoni2016}.

To quantify the scale of the nebulae, here we computed the area of the extended \lya\ emission as the sum of pixels above a given \lya\ SB level, which is less sensitive to possible asymmetries of the nebulae. To define a common SB value for both the MUDF and QSO MUSEUM datasets, we considered the \rev{shallowest} MUSE datacube (i.e. ID21), \rev{finding a SB of $1.2\times10^{-18}$ erg s$^{-1}$ cm$^{-2}$ arcsec$^{-2}$ (at the $2\sigma$ statistical level) at $z=3.218$}, which we consider as a reference (${\rm SB}_{\rm ref}$). We then corrected this value for the cosmological dimming as ${\rm SB}_{\rm cut} = {\rm SB}_{\rm ref}\times(1+z_{\rm ref})^4/(1+z)^4$.
For the bright MUDF quasar, \qsoone, we obtain $\sim$114 arcsec$^2$, which translates into 6732 proper kpc$^2$ (with a conversion of $\simeq7.7$ kpc per arcsec at $z=3.22$). For \qsotwo, we obtain %1134 pixels 
$\sim$45 arcsec$^2$, or equivalently 2679 proper kpc$^2$. This value may be slightly underestimated because of the presence of a bright star to the South of the nebula.
\rev{Figures~\ref{fig:areax} and \ref{fig:area} present the area of the extended \lya\ emission as a function of the nuclear luminosity at rest-frame 2 keV and 2500 \AA.} We do not observe any statistically significant correlation of the area with either of these two luminosities: small and large nebulae are equally distributed within a similar range of nuclear ionising powers. The absence of a trend between area and nuclear luminosity is also confirmed if we exclude the X-ray weak and the two radio bright quasars. \rev{The area of the nebulae is also not correlated with the X-ray photon index, as shown in the upper right panel of Figure~\ref{fig:areax}}. 

For completeness, we also considered whether there is any trend between the \lya\ emission of the nebulae ($L_{\rm Ly\alpha}$) and the quasar rest-frame 2 keV luminosity, but we found none (Figure~\ref{fig:areax}). The luminosity values for the extended \lya\ emission have been taken directly from the relevant papers (i.e., Table~1 in \citealt{lusso19} and Table~2 in \citealt{arrigoni2019a}). For the MUDF data, even if the field has now reached a much deeper MUSE exposure than that considered in the original work, the extended emission at the edge of the SB tails is negligible with respect to the total one. Therefore, the values for $L_{\rm Ly\alpha}$ have not changed significantly even considering the deeper data. 
The $L_{\rm Ly\alpha}$ values for the QSO MUSEUM sample have been computed by employing much shallower data instead. For the quasar in common between the MUDF and QSO MUSEUM (ID22, i.e., \qsoone), the $L_{\rm Ly\alpha}$ value in the latter analysis is higher by a factor of $\sim1.5$ (in flux) than the one reported in \citet{lusso19}. This 0.2 dex difference is likely caused by the fact that, to identify the \lya\ extended emission around the quasars, in the QSO MUSEUM analysis a minimum ``volume'' of 1000 connected voxels above S/N$>$\,2 was considered, against 2500 connected voxels at S/N\,$\geq$\,2.5 in \citet{lusso19}. This implies that some noise could be included in the QSO MUSEUM flux measurements, but changing the thresholds would just introduce a systematic shift that does not change the main result, that is the absence of a trend between $L_{\rm Ly\alpha}$ and $\lx$.
% discussion of Llya-area relation
\rev{Finally, Figure~\ref{fig:area} also shows the $L_{\rm Ly\alpha}$ integrated emission of the nebulae as a function of their area. These two quantities are correlated, with the area increasing with increasing \lya\ emission, in broad agreement with what found by \citet{mackenzie2021}. The best fit regression line is} 
\begin{multline}
%\begin{equation}
    \log (L_{\rm Ly\alpha}/[10^{44} \rm erg\,s^{-1}]) = (1.16\pm0.06) \log (\rm area/[pkpc^2]) + \\(-4.42\pm0.77)
%\end{equation}
\end{multline}
\rev{with a dispersion of 0.12 dex. The observed relation is in agreement with what found by \citet[][see their Table~1 for the resulting relationships with the physical area]{ab2023}, who argue that host-galaxy inclination could be the main driver for the shape of the nebulae and their brightness, and thus for the relation itself (see their Section~5). 
}

Overall, \rev{our results are qualitatively at odds with the interpretation by \cite{mackenzie2021} that the quasar ionising power is the main driver of the nebulae properties (i.e. luminosity and size; see also \citealt{rz2013})}. Yet, given all the caveats discussed above, we caution that a one-to-one comparison of the sizes and areas with the luminosities measured in MUDF and QSO MUSEUM datacubes and other studies is not straightforward, because of the different sensitivity limits, techniques, and definition of the extent of the diffuse \lya\ emission. Additionally, the geometry of the host galaxy may also play a role (edge-on versus face-on), giving rise to different nebular morphologies \citep[e.g.][]{costa2022,ab2023,rz2013}. 
\rev{The absence of any correlation between the quasar and the nebular properties (if confirmed with larger samples), combined with the fact that there is a tight correlation between the emission of the nebulae and their size, imply that the main drivers %for the physical mechanism that control the morphology 
the extent of the nebulae could be host galaxy inclination and the physical properties of the environment, such as its density and radiation (assuming the emission is due to photoionization and perhaps scattering)}.  Whilst the quasars provide the main reservoir of ionising photons (see the discussion in Section 4.1 in \citealt{mackenzie2021}) that ultimately powers the nebulae, our results suggest that the actual extent of the diffuse emission and its morphology (e.g. asymmetries) seem to be the result of scattering processes in the CGM \citep[e.g.][]{arrigoni2019a,costa2022}.
As the fraction of X-ray emission to optical/UV is less than 1\% for both \qsoone and \qsotwo (see appendix~\ref{Spectral energy distributions}), this implies that the X-ray emission is not contributing significantly to the overall photon budget. We caution, however, that this fraction is computed by assuming that the X-ray emission that illuminates the \lya\ nebulae of the MUDF quasars is the same as observed.  

\section{Conclusions}
\label{conclusions}
We present \xmm observations of the MUSE Ultra Deep Field (MUDF), a unique region of the sky that hosts two quasars at $z\simeq3.22$ with close separation ($\sim500$ kpc). Observations at high energy characterise the innermost region of these quasars with physically associated extended \lya\ nebulae, and provide imaging the MUDF and the environment of this pair at much larger scales %($>1^\prime$ field of view of MUSE). 
than those covered by either MUSE or \hst. 
Thus, MUSE+\xmm observations represent the first view of the assembly of a potentially massive $z\simeq3.2$ overdensity. We searched for additional X-ray detections within the F140W/\hst\ field of view around the MUDF. We extended this search to a distance of 10$^{\prime}$ to provide a detection list that roughly matches the distance of the third $z\sim3.2$ quasar \qsothree. We find 119 X-ray detections, with seven detections inside the MUDF/\hst field. The three already known quasars (i.e., \qsoone, \qsotwo, and a lower redshift quasar in close spatial proximity to a foreground $z\simeq0.67$ galaxy group) within the MUSE coverage are all detected. These sources are the only ones detected within the MUSE footprint. For the additional sources in the \hst field, only one has an ambiguous counterpart (\textsc{src\_num}=159), for which a sub-arcsecond accurate X-ray position is required. Given the richness of the \hst field, the \chandra sub-arcsecond spatial resolution and low background are essential to further disentangle multiple sources that are blurred in the large \xmm point-spread function.

\rev{We find that both quasars in the MUDF are X-ray weak, outliers of the $\lx-\lo$ relation, with a \ion{C}{iv} line emission in agreement with other X-ray weak quasars in the literature at $z\simeq3$ and matching UV luminosities.} This result is compatible with the higher probability to observe an X-ray weak quasar at these redshifts, as suggested by previous works analysing quasar samples at similar redshift and UV luminosities \citep[e.g.][]{nardini2019,zappacosta2020}, and it might be interpreted in a starved X-ray corona scheme associated with an ongoing wind phase. If the wind starts off in the vicinity of the super massive black hole, the accretion rate -- and thus the UV light that reaches the X-ray corona -- will diminish, depriving the hot coronal plasma of seed photons and resulting in an X-ray weak quasar. \rev{Yet, at high UV luminosities there will still be sufficient ionising photons to produce a strong \ion{C}{iv} line emission as detected in X-ray weak quasars, which have higher $L_{\rm C\,IV}$ compared to normal quasars at similar X-ray luminosities \citep[][]{lusso2021,trefoloni2023}.}

We do not observe any trend between the area of the \lya\ nebulae and nuclear luminosities at either the rest-frame 2 keV or 2500 \AA. The area of the nebulae does not correlated with the X-ray photon index nor with the integrated band flux in the hard band (2--10 keV). Quasars with similar luminosities can have very diverse \lya\ areas. The absence of a relation between the extent of the nebula and the quasar ionising power is at variance with what is observed by \cite{mackenzie2021}, who found a correlation between the size of the nebula (defined as the maximum extent) and the luminosity of the quasar in both UV continuum and \lya. We also do not find any statistically significant trend between $L_{\rm Ly\alpha}$ and $\lx$.
\rev{Finally, the MUDF quasars are in agreement with the relation between the $L_{\rm Ly\alpha}$ integrated emission of the nebulae and their area recently published by \citet{ab2023}, suggesting that host-galaxy inclination could be amongst the main drivers for the morphology and brightness of the nebulae.}
However, a direct comparison of the sizes and areas with luminosities between different studies is challenging, due to e.g. different sample statistics, sensitivity limits, techniques, and definition of the extent of the diffuse \lya\ emission). 
Our findings, if confirmed with larger samples, suggest that the nebular morphology is mainly driven by the physical properties of the environment in the host galaxy (e.g. geometry, density, temperature, column density, and filling factor or clumpiness), rather than the quasar power.

\section*{Acknowledgements}
We thank the anonymous reviewer for their thorough reading and for useful comments and suggestions that have improved the clarity of the paper.
E.L. acknowledges the support by the Fondazione Cassa di Risparmio Firenze (grant No 45780). 
S.C. acknowledges financial support of the Italian Ministry of Education, University, and Research with PRIN 201278X4FL and the ``Progetti Premiali'' funding scheme.
P.D. acknowledges support from the NWO grant 016.VIDI.189.162 (``ODIN") and from the European Commission's and University of Groningen's CO-FUND Rosalind Franklin program.
Based on observations with the NASA/ESA Hubble Space Telescope obtained from the MAST Data Archive at the Space Telescope Science Institute, which is operated by the Association of Universities for Research in Astronomy, Incorporated, under NASA contract NAS5-26555. Support for program numbers 15637 and 15968 was provided through a grant from the STScI under NASA contract NAS5-26555. These observations are associated with program numbers 6631, 15637, and 15968. 
The MUSE portion of this project has received funding from the European Research Council (ERC) under the European Union’s Horizon 2020 research and innovation programme (grant agreement No 757535) and by Fondazione Cariplo (grant No 2018-2329).

%%%%%%%%%%%%%%%%%%%%%%%%%%%%%%%%%%%%%%%%%%%%%%%%%%
\section*{Data Availability}
%The inclusion of a Data Availability Statement is a requirement for articles published in MNRAS. Data Availability Statements provide a standardised format for readers to understand the availability of data underlying the research results described in the article. The statement may refer to original data generated in the course of the study or to third-party data analysed in the article. The statement should describe and provide means of access, where possible, by linking to the data or providing the required accession numbers for the relevant databases or DOIs.
Raw X-ray data are available via the \xmm\ Science Archive (XSA). The processed \hst\ and \xmm\ observations are available at \url{https://archive.stsci.edu/hlsp/mudf: (doi:10.17909/81fp-2g44)}.

%%%%%%%%%%%%%%%%%%%% REFERENCES %%%%%%%%%%%%%%%%%%

% The best way to enter references is to use BibTeX:

\bibliographystyle{mnras}
\bibliography{bibl} % if your bibtex file is called example.bib

% Alternatively you could enter them by hand, like this:
% This method is tedious and prone to error if you have lots of references
%\begin{thebibliography}{99}
%\bibitem[\protect\citeauthoryear{Author}{2012}]{Author2012}
%Author A.~N., 2013, Journal of Improbable Astronomy, 1, 1
%\bibitem[\protect\citeauthoryear{Others}{2013}]{Others2013}
%Others S., 2012, Journal of Interesting Stuff, 17, 198
%\end{thebibliography}

%%%%%%%%%%%%%%%%%%%%%%%%%%%%%%%%%%%%%%%%%%%%%%%%%%

%%%%%%%%%%%%%%%%% APPENDICES %%%%%%%%%%%%%%%%%%%%%

\appendix

\section{X-ray observations of QSO MUSEUM}
\label{app.A}
% -----------------
\begin{table*}
\centering
\caption{Properties of the QSO MUSEUM sources with X-ray observations. } \label{tabA1}
\begin{tabular}{lcccccccc}
\hline
\hline
Object & $z_{\rm sys}^{\mathrm{a}}$ & ID & $C/\nu$ & $\Gamma_{\rm X}$ & $\log (\nu F_\nu)_{2\,{\rm keV}}$$^{\mathrm{b}}$ & $L_{\rm UV}^{\mathrm{c}}$ & $(\nu L_{\nu})_{\rm CIV}^{\mathrm{d}}$ &Area\\          
&  &  &  &  & erg\,s$^{-1}$\,cm$^{-2}$ & erg\,s$^{-1}$\,Hz$^{-1}$ & erg\,s$^{-1}$ & pkpc$^2$ \\   \hline  
QSO\,B0114$-$0857         & 3.204 & 31 & 55/65     &  $1.737\,(1.468,2.011)$ & $-13.96\,(-14.04,-13.87)$ &  $1.33\pm0.05$ &  $1973.0\pm76.2$  &  2767\\      
QSO\,B0537$-$286          & 3.139 & 49 & 5757/6344 &  $1.274\,(1.266,1.282)$ & $-12.48\,(-12.48,-12.47)$ &  $0.28\pm0.07$ &  $536.6\pm16.3$   &  5732\\      
                          &       &    & 4598/4639 &  $1.157\,(1.150,1.163)$ & $-12.35\,(-12.35,-12.34)$ &                &                   &      \\      
                          &       &    & 404/425   &  $1.236\,(1.208,1.264)$ & $-12.76\,(-12.78,-12.74)$ &                &                   &      \\      
SDSS\,J0947+1421 & 3.039 & 6  & 643/680   &  $2.325\,(1.921,2.796)$ & $-13.56\,(-13.58,-13.53)$ &  $1.59\pm0.04$ &  $3451.7\pm121.5$ &  3467\\      
QSO\,J0958+1202           & 3.306 & 9  & 8/21      &  $1.640\,(1.513,1.765)$ & $-13.26\,(-13.48,-13.07)$ &  $1.33\pm0.05$ &  $3151.6\pm105.7$ &  4784\\      
7C\,1013+2053             & 3.108 & 54 & 113/132   &  $1.736\,(1.555,1.921)$ & $-13.12\,(-13.24,-12.10)$ &  $0.26\pm0.06$ &  $ 984.1\pm18.7$  &  4675\\      
LBQS\,1244+1129           & 3.155 & 59 & 449/504   &                         & $-13.86\,(-13.90,-13.82)$ &  $0.85\pm0.05$ &  $1849.0\pm48.9$  &  6663\\      
CTQ\,669                  & 3.219 & 21 & 129/137   &  $2.513\,(1.718,3.388)$ & $-13.34\,(-13.66,-13.11)$ &  $0.92\pm0.05$ &  $1797.4\pm58.9$  &  7364\\      
QSO\,B2348$-$404          & 3.332 & 27 & 11/15     &                         & $-13.59\,(-15.84,-15.40)$ &  $1.11\pm0.07$ &  $1828.7\pm73.4$  &  6265\\                   
\hline
\end{tabular}
 \flushleft\begin{list}{}
 \item {\it Notes.} Values within the parenthesis represent the minimum and maximum value of the correspondent parameter. ${}^{\mathrm{a}}${ Redshifts are obtained from the peak of the \lya\ emission in the quasar spectrum (from \citealt{arrigoni2019a}). ${}^{\mathrm{b}}$ Flux at rest-frame 2 keV. ${}^{\mathrm{c}}$ Monochromatic continuum luminosity at rest-frame 2500\,\AA\ normalised to $10^{32}$ erg\,s$^{-1}$\,Hz$^{-1}$. ${}^{\mathrm{d}}$ Integrated \ion{C}{iv} luminosity normalised to $10^{42}$ $\rm erg~s^{-1}$.}
 \end{list}
\end{table*}
% -----------------

As mentioned in the main text, neglecting \qsoone and \qsothree (ID22 and ID23, respectively), other eight out of the 61 sources in the QSO MUSEUM sample \citep{arrigoni2019a} have been observed in the X-rays by \xmm and/or \chandra. We list below the observations that have been considered in this work. For each target, the key properties relevant to the present discussion, including those inferred from the X-ray analysis, are summarized in Table~\ref{tabA1}. \\
$\bullet$~QSO\,B0114$-$0857 was serendipitously observed by \chandra on 2020 October 16 and two days later, for a cumulative exposure of 37.7 ks. The source lies about 3 arcmin off-axis, and the spectra were extracted from a circular region with radius of 5 arcsec. The analysis was carried out simultaneously on the two data sets, and the spectral fits were performed over the 0.6--6 keV band, where 74\,$(\pm9)$ net counts are collected. \\
$\bullet$~QSO\,B0537$-$286 was targeted several times, first by \xmm on 2000 March 19, with net exposures of 19.4 ks for the pn and 38.8 ks for both MOS cameras. Another observation was performed on 2005 March 20. Neglecting a short snapshot on the same day, the on-source times are 31.8 ks, 51.0 ks, and 49.7 ks for pn, MOS1, and MOS2, respectively. The spectra were extracted from circular regions with radii of 35 (30) arcsec in the first (second) epoch for all the detectors. No exposure is affected by significant background flares, and about 31 and 45 kilo-counts are available over the entire 0.3--10 keV EPIC band. In between the two \xmm visits, QSO\,B0537$-$286 was also observed by \chandra on 2003 July 21 for 30 ks. The extraction region is 8 arcsec wide as the source is $>$\,5 arcmin away from the nominal aimpoint and PSF distortion effects are non negligible. The spectral analysis was performed over the 0.5--7 keV band, where $\sim$\,3600 counts are collected with a 0.1 per cent background level. In order to take into account the source variability, we considered in the plot the average flux of these three observations, whilst the error covers the min-to-max flux range. \\
$\bullet$~SDSS\,J094734.19+142116.9 is part of the sample of luminous blue quasars at $z\simeq3$ with pointed \xmm observations discussed by \citet{nardini2019}. The source was observed on 2017 April 28, with good-time exposures of 24.2 ks (pn), 30.1 ks (MOS1), and 30.0 ks (MOS2). Over 1,000 net counts were cumulatively obtained by the three EPIC detectors. The relevant X-ray properties have been directly retrieved from the spectral analysis of \citet{nardini2019}. An earlier \chandra snapshot (1.6 ks) provides much looser constraints and is neglected.  \\
$\bullet$~QSO\,J0958+1202 can instead rely only on one such \chandra observation, performed on 2012 April 22. Despite the short exposure (1.6 ks), the source is robustly detected, with 21 counts at 0.5--7 keV within a radius of 3 arcsec (estimated background 0.06), allowing a basic spectral analysis. \\
$\bullet$~7C\,1013+2053 was observed by \chandra on 2018 January 24 for 10 ks. The spectrum was extracted from a circular region with radius of 3 arcsec, and the source and estimated background counts in the 0.5--7 keV band are, respectively, 205 and $\sim$\,0.4. \\
$\bullet$~LBQS\,1244+1129 is also included in the study by \citet{nardini2019}, to which we refer for the details on the spectral analysis. The \xmm observation took place on 2017 July 03, with exposures of 32.6 ks for the pn and 38.5 ks for both MOS cameras. About 560 source counts were collected at 0.5--8 keV. \\
$\bullet$~CTQ\,669 was serendipitously observed by \xmm on 2011 October 11, with exposures of 8.8 ks for the pn and 11.5 ks for both MOS detectors. The spectra were extracted from circular regions with radius of 20 arcsec, with 112\,$(\pm14)$ source counts at 0.3--8 keV. The pn spectrum, however, turned out to be inconsistent with the two MOS ones in terms of intensity, being brighter by about a factor of 3. A similar effect is noticed also in the relative fluxes of the three EPIC instruments reported in the latest \xmm catalogue \citep[4XMM--DR12;][]{webb2020} of serendipitous X-ray sources, and its origin is unclear. The observation is not affected by background flares, so a possible explanation could be the proximity of the target to the edge of the chip in the pn field of view, which might result in some problem with the effective-area correction in the auxiliary response file. We conservatively assumed the average flux of the pn and MOS observations, whilst the error is the half-difference between the two values. \\
$\bullet$~QSO\,B2348$-$404 was serendipitously observed by \xmm on 2017 May 14, and is marginally detected in 4.4 ks only with the pn with $\sim$\,15 counts at 0.3--5 keV. The source is not detected with MOS1, and falls outside the MOS2 field of view.

% --------
\begin{figure*}
	\includegraphics[width=0.9\textwidth]{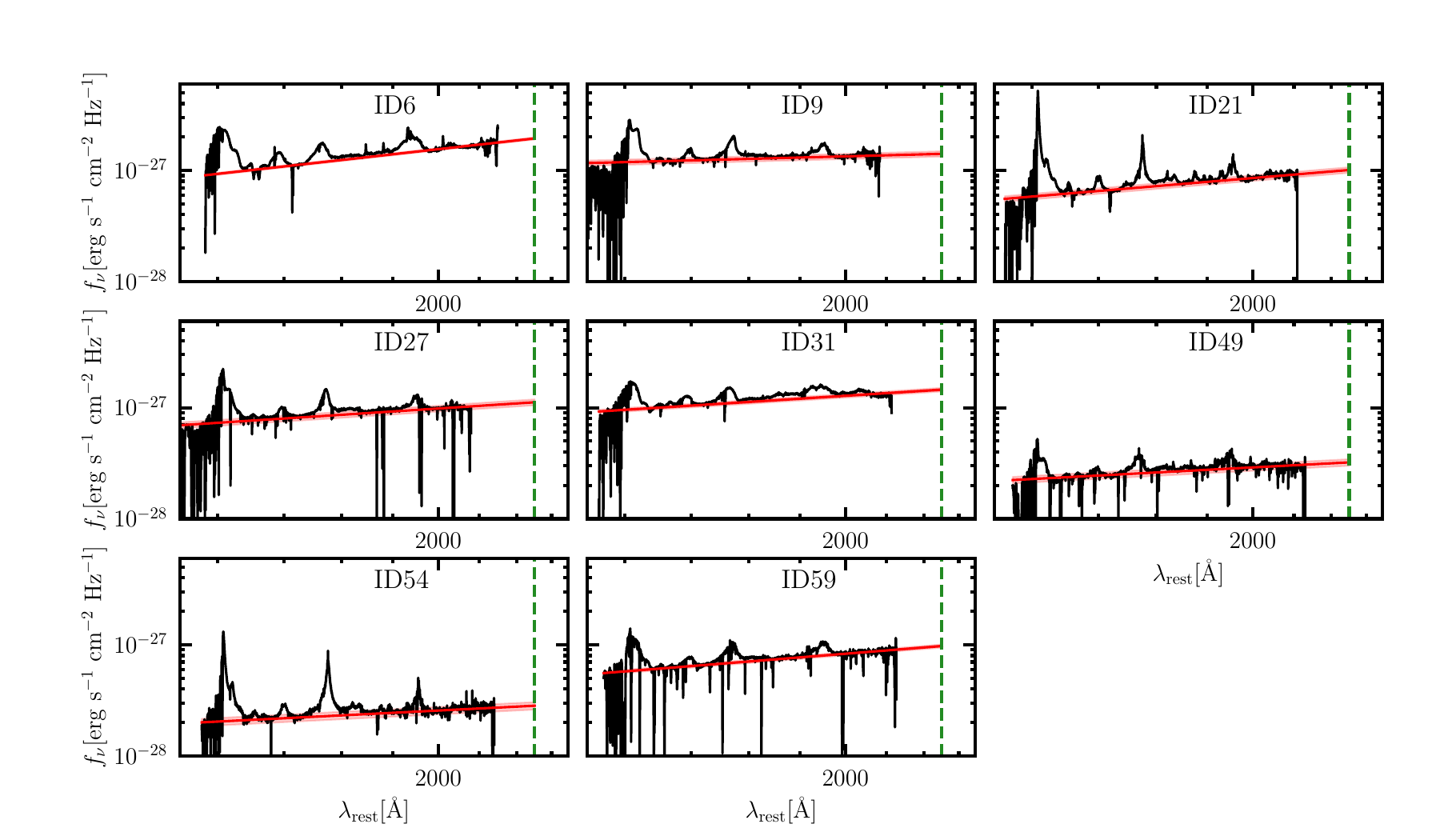}
    \caption{MUSE spectra of the eight quasars with X-ray archival data in the QSO MUSEUM sample. The red solid line represents the continuum (along with uncertainties in light red). The dashed line marks the 2500 \AA\ wavelength at rest. All panels have the same ranges on both axes to ease comparison. The source ID (as listed in \citealt{arrigoni2019a}) is shown at the top of each panel.}
    \label{fig:spectra-museum}
\end{figure*}
% --------
% --------
\begin{figure*}
	\includegraphics[width=0.9\textwidth]{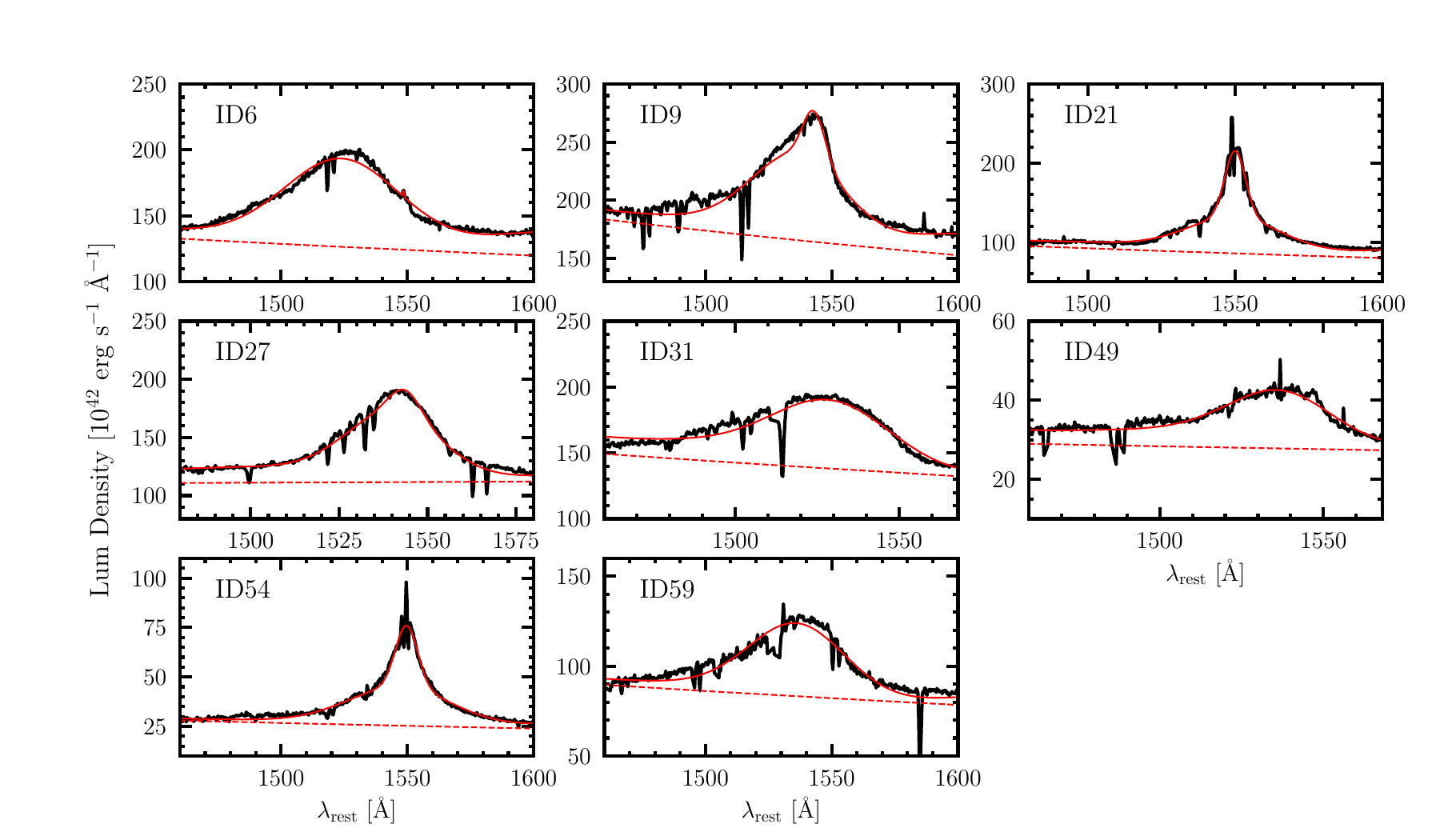}
    \caption{Fits to the \ion{C}{iv} emission line of the 8 MUSE quasar spectra in the QSO MUSEUM sample with X-ray archival data. The spectra are shown with a black line, the best-fitting model in red, and the continuum with the red dashed line.}
    \label{fig:civ-museum}
\end{figure*}
% --------
\section{Ultraviolet observations of QSO MUSEUM}
\label{app.B}
Figures~\ref{fig:spectra-museum} and \ref{fig:civ-museum} present the fit of the MUSE spectra of the eight quasars with X-ray archival data in the QSO MUSEUM sample. The flux, along with its uncertainty, at rest-frame 2500 \AA\ has been obtained from the extrapolated best-fit continuum. 

% spectral energy distribution ------------

\section{Spectral energy distributions of the MUDF $z\simeq3$ quasars}
\label{Spectral energy distributions}
% --------
\begin{figure*}
	\includegraphics[width=0.48\textwidth]{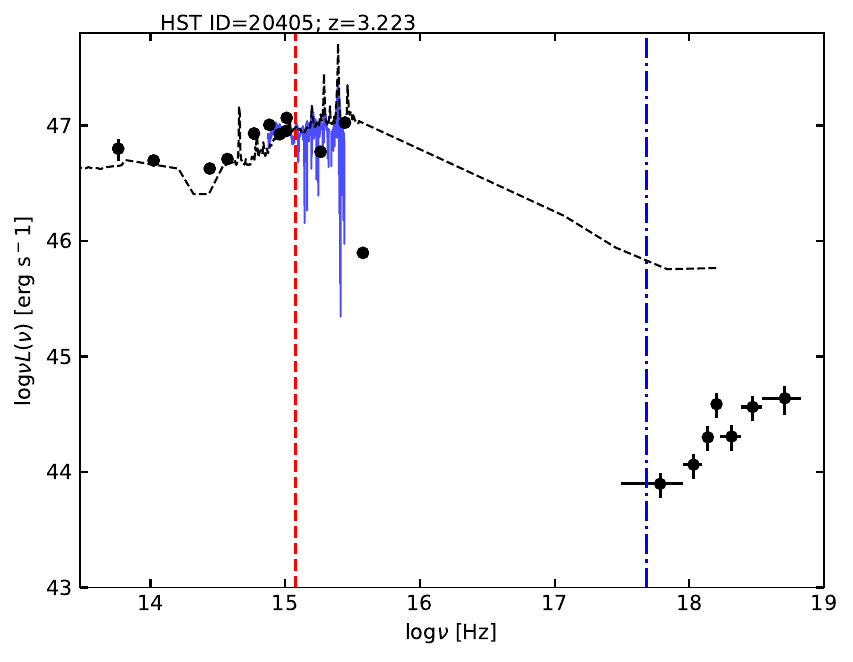}
	\includegraphics[width=0.48\textwidth]{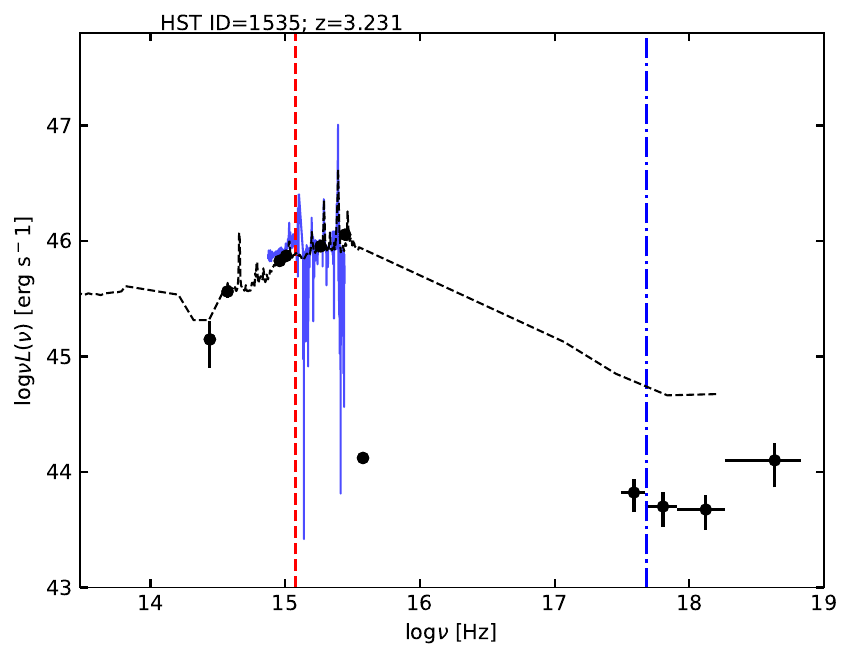}
    \caption{Rest-frame broad-band photometric spectral energy distributions for \qsoone\ (\hst ID=20405) and \qsotwo\ (\hst ID=1535) from the near-infrared to the X-rays. Data are shown with black points. The MUSE plus \hst spectrum for the two quasars is shown with a blue line. We also overplotted the SED for radio-quiet AGN published by \citet{shang2011} as a reference. The vertical red dashed and blue dot-dashed lines mark the 2500 \AA\ and the 2 keV energies, respectively.}
    \label{fig:seds}
\end{figure*}
% --------

For completeness, we also compiled the broadband photometric SEDs, from the near-infrared to the X-rays, for the two MUDF quasars (Figure~\ref{fig:seds}). 
\qsoone has a rather good photometric coverage, from the rest-frame near-infrared with WISE (with a S/N\,$>$\,3) and 2MASS, to the optical/UV with \hst\ (F140W, F125W, F702W, F450W, F336W, see \citealt{revalski2023} for details), for a total of 12 photometric data points. \qsotwo is instead  detected in WISE/W1 and W2 (with a S/N\,$>$\,2) only, but not in 2MASS. Together with the \hst\ photometry, the rest-frame near-infrared to optical/UV SED has 7 data points. The X-ray data are also included, which have been corrected for Galactic absorption. Following a simplified approach as in \citet[][see their Section 4.2]{lusso2013}, we computed the bolometric luminosity as the integrated emission from 1 $\mu$m up to 1 keV\footnote{The monochromatic luminosity at the rest-frame energy of 1 keV is about $3\times10^{45}$ erg s$^{-1}$ for both quasars.}, as the quasar emission below and above this range is considered to be reprocessed. We find $L_{\rm bol}\simeq2\times10^{47}$ and $2\times10^{46}$ erg s$^{-1}$ for \qsoone and \qsotwo, respectively. As the rest-frame 2--10 keV luminosity is $2.3\times10^{44}$ and $7.2\times10^{43}$ for the brighter and the fainter quasar in MUDF, we obtain an X-ray-to-optical/UV ratio less than 1\% for both sources, thus implying that the X-ray emission is a negligible fraction of the total one for the two MUDF quasars. 
%------------------------

%If you want to present additional material which would interrupt the flow of the main paper, it can be placed in an Appendix which appears after the list of references.

%%%%%%%%%%%%%%%%%%%%%%%%%%%%%%%%%%%%%%%%%%%%%%%%%%

% Don't change these lines
\bsp	% typesetting comment
\label{lastpage}
\end{document}